%% file: main.tex
\documentclass[11pt]{article}

\usepackage[preprint]{acl}

\usepackage{times}
\usepackage{latexsym}
\usepackage{amssymb}
\usepackage{amsmath}
\usepackage{booktabs}
\usepackage{enumitem}
\usepackage{subcaption}
\usepackage{comment}
\usepackage{rotating}
\usepackage{tikz}
\usepackage{colortbl}
\usepackage{tabularx} % in preamble
\usepackage{tikz}
\usepackage{pgfplots}
\pgfplotsset{compat=1.18}
\usepgfplotslibrary{groupplots}
\usepackage{colortbl}
\usetikzlibrary{shapes.geometric}
\usepackage{multirow}

\definecolor{mpblue}{HTML}{1F77B4}
\definecolor{mporange}{HTML}{FF7F0E}
\definecolor{mpgreen}{HTML}{2CA02C}
\definecolor{mpred}{HTML}{D62728}
\definecolor{mppurple}{HTML}{9467BD}
\definecolor{mpbrown}{HTML}{8C564B}
\usetikzlibrary{shapes.geometric}
\definecolor{agentblue}{RGB}{55,138,221}
\definecolor{agentred}{RGB}{226,75,74}
\definecolor{panelbg}{RGB}{245,245,248}

\usepackage{tcolorbox}
\tcbuselibrary{skins,breakable}

\definecolor{promptbg}{HTML}{F7F7F8}
\definecolor{promptframe}{HTML}{4A90D9}
\definecolor{headerblue}{HTML}{4A90D9}
\definecolor{fieldbg}{HTML}{E8E8E8}
\definecolor{outputbg}{HTML}{F0F0F0}

\tcbset{
prompt box/.style={
  enhanced, breakable, colback=promptbg, colframe=promptframe,
  fontupper=\small, left=4mm, right=4mm, top=4mm, bottom=3mm,
  arc=1mm, boxrule=0.5pt
},
header box/.style={
  colback=headerblue, colframe=headerblue, arc=1mm, boxrule=0pt
},
field header/.style={
  colback=fieldbg, colframe=fieldbg, arc=0.5mm, boxrule=0pt,
  fontupper=\small\bfseries, left=2mm, right=2mm,
  top=0.5mm, bottom=0.5mm
},
output box/.style={
  colback=outputbg, colframe=outputbg, arc=0.5mm, boxrule=0pt,
  fontupper=\small\ttfamily, left=2mm, right=2mm,
  top=1mm, bottom=1mm
},
}

\newcommand{\pvar}[1]{\texttt{\{#1\}}}

\usepackage[T1]{fontenc}
\usepackage[utf8]{inputenc}

\usepackage{microtype}

\usepackage{inconsolata}

\usepackage{graphicx}

\title{When is Routing Meaningful? \\Diversity and Robustness in Language Model Societies}

\def\authorsep{\unskip\enspace\enspace\enspace\enspace}

\author{Fantine Huot \authorsep Michael Kaisers \authorsep Mirella Lapata \vspace{5pt} \\ 
        Google DeepMind \\ \texttt{\{fantinehuot,mkaisers,lapata\}@google.com}} 

\begin{document}
\maketitle
\begin{abstract}
Routing policies for multi-model systems are evaluated almost
exclusively on task accuracy and inference cost. We argue that
two properties, orthogonal to performance, determine whether
routing is meaningful. First, the society of actors must be
behaviourally differentiated: if all actors respond identically,
routing is vacuous. Second, the routing policy must be stable: surface-form variants
of a query should be assigned to the same actor. High task
accuracy is compatible with violating both properties, since a
router can operate over a redundant society or assign queries
inconsistently, preventing specialisation regardless of
performance.  We adapt Hierarchic Social
Entropy (HSE) to language-model societies and introduce a
perturbation-based robustness metric to diagnose these failure
modes. Applied to EmbedLLM and RouterBench, we find that HSE
exhibits strong diminishing returns, suggesting that a curated
subset of fewer than ten agents recovers most available diversity
in a large pool --- a practical coreset heuristic for society
design. We further find that KNN routers gain accuracy from
specialist societies but collapse in robustness under
perturbation, while prompted routing remains stable across all
perturbation types --- illustrating that accuracy and
meaningfulness can sharply diverge.

\end{abstract}

\section{Introduction}
\label{sec:introduction}

Routing is becoming a central mechanism for coordinating systems
composed of multiple language models, tools, or agents
\citep{wu2024autogen, hong2024metagpt, li2023camel, yue2025masrouter}.
Existing work on LLM routing largely evaluates routers by downstream
utility: whether they improve task accuracy, reduce cost, or select the
best-performing model for a query
\citep{chen2024frugalgpt, ong2025routellm, hu2024routerbench, jiang2023llmblender}.
 A router is not merely a
predictor of task performance; it is also the coordination mechanism that determines how work is distributed across a society of actors. Its usefulness, therefore, depends on two structural properties. First, the society must contain behaviourally differentiated actors: if all actors respond in the same way, routing is vacuous
\citep{balch2000hierarchic, bettini2025system}. Second,  the routing policy should be robust to superficial variation:
semantically-equivalent variants should not be sent to different actors
merely because of changes in spelling, syntax, or wording
\citep{sclar2024quantifying}. Without such stability, actors cannot reliably specialise.

To characterise the space of choices available to a router, we adapt Hierarchic Social Entropy \citep{balch2000hierarchic} to language-model societies, measuring behavioural diversity from model outputs rather than parameters \citep{stanley2019designing}.
To characterise the stability of the router's choices, we introduce a perturbation-based robustness metric that measures whether semantically related inputs are assigned to the same actor.

We validate HSE on two benchmarks, EmbedLLM \citep{zhuang2025embedllm} and
RouterBench \citep{hu2024routerbench}, comparing a default society of real-world
language models against  synthetic societies of
purpose-designed experts varying in role and
expertise overlap. 

We find that specialist societies achieve substantially
higher HSE than pools of real-world models of equivalent
size, suggesting that behavioural diversity in deployed
model pools is lower than commonly assumed.
We further observe strong diminishing returns: fewer than ten
agents suffice to capture most available diversity in
EmbedLLM, and four agents in RouterBench, providing a
practical heuristic for society design. Interestingly, higher HSE does not imply higher robustness: KNN routers achieve their best accuracy on specialist societies but their worst robustness, while prompted routing remains stable across all HSE levels and perturbation types.
Our contributions are summarised as follows: 

\begin{itemize}[itemsep=0pt,labelindent=0pt,leftmargin=*]
\item We introduce the first behavioural diversity metric
for LM routing analysis, adapting Hierarchic Social
Entropy to language-model societies to characterise
when routing is meaningful. 

\item  We introduce a
perturbation-based metric that quantifies how stably a
routing policy assigns semantically-equivalent queries to
the same actor, paired with a five-level perturbation
taxonomy for fine-grained analysis by perturbation
type. 

\item 
We show that HSE exhibits strong diminishing returns with
society size: a curated subset of fewer than ten agents
recovers most available diversity in a large real-world pool,
providing a practical coreset heuristic for society design.

\end{itemize}
\section{Related Work}
\label{sec:related_work}

\paragraph{LLM Routing and Model Selection}
A growing body of work studies routing as a mechanism for selecting among language models with different capabilities, costs, and latency profiles. In this setting, the router is typically evaluated by its ability to improve the trade-off between task performance and inference cost. FrugalGPT~\citep{chen2024frugalgpt}, for example, formulates
LLM use as a cost-sensitive cascade, where cheaper models are queried before more expensive ones in order to reduce inference cost while maintaining task accuracy. RouterBench~\citep{hu2024routerbench}
provides a benchmark for multi-LLM routing, evaluating how well routing policies choose among alternative models across tasks and model families. RouteLLM~\citep{ong2025routellm} learns routing policies from preference data, routing queries between stronger and weaker models in order to preserve response quality while reducing the use of expensive
models.

More recently,
MasRouter~\citep{yue2025masrouter} extends routing to multi-agent settings, jointly optimising collaboration mode, role allocation, and model selection, building on multi-agent LLM frameworks in which agent
roles and interaction protocols are central design
choices~\citep{wu2024autogen, hong2024metagpt, li2023camel}.

\paragraph{Diversity Metrics for Agent Systems} The measurement of behavioural diversity in multi-agent systems has been studied primarily in robotics. \citet{balch2000hierarchic} introduces Hierarchic Social Entropy (HSE) as an information-theoretic measure of robot group diversity that captures both the number and relative size of behavioural clusters across all taxonomic scales. More recently, \citet{bettini2025system} propose System Neural Diversity (SND), which uses Wasserstein distance over multivariate robot-policy action distributions. SND requires all the agents to compute probabilities on the same action space and would not translate to heterogeneous LM agents with access to different tools. Our work adapts HSE to LM agent societies, replacing Euclidean inter-agent distance with cosine similarity over behavioural vectors, and is the first to apply such metrics to LM routing analysis. The ensemble learning literature has also studied diversity as a design objective~\cite{NIPS1994_b8c37e33}, and quality-diversity optimisation~\cite{mouret2015mapelites} explicitly searches for a diverse set of high-performing solutions. 

\paragraph{Specialisation in Multi-Agent Systems}
A related line of work studies how individual actors come to
occupy stable, distinct roles within a group — a prerequisite
for routing to be meaningful over time.
In multi-robot systems, specialisation is typically measured
by the stability of roles rather than by task accuracy.
\citet{labella2006division} introduces response threshold
models in which an agent's propensity to perform a task
decreases each time it is performed by another agent;
specialisation is said to emerge when an agent's threshold
for a given task remains persistently low.
\citet{nitschke2008emergent} operationalises this differently
via Collective Neuro-Evolution (CONE), defining a specialised
agent as one that assumes a specific behavioural role for more
than 50\% of its operational lifetime.

Our routing robustness metric formalises this intuition for
LM societies, measuring whether semantically-equivalent
query variants are stably assigned to the same actor —
the distributional consistency that these robotics accounts
identify as the precondition for specialisation.

\paragraph{Robustness to Prompt Perturbations}
A separate line of work studies the robustness of language models under input perturbations, including character-level noise, lexical substitutions, syntactic transformations, and paraphrases. A common finding is that meaning-preserving changes can cause large accuracy swings  \cite{sclar2024quantifying}  across model sizes and instruction tuning. Prior work typically responds by training models to produce consistent outputs across clean and perturbed prompts~\citep{qiang2024prompt}, or by recommending that evaluations report performance ranges rather than point estimates~\citep{sclar2024quantifying}.
We study robustness at a different level: not whether the final answer is stable under perturbation, but whether the \emph{routing decision} is. 

By measuring whether perturbed queries are routed to the same actor as the original, we evaluate whether the routing policy induces stable task distributions over actors.

%%%%%%%%%%%%%%%%%%%%%%%%%%%%%%%%%%%%%%%%%%%%%%%%%%%%%%%%%%%%%
\section{Preliminaries}
\label{sec:preliminaries}

We introduce the notation  used throughout the paper and  state the two structural properties we study.

Let $\mathcal{R} = \{r_1, r_2, \dots, r_N\}$ be a \emph{society}
of $N$ actors, where each actor may correspond to a distinct model, prompt, tool-using agent, or role-specialised configuration.  Actors may differ in architecture, scale, training data, or system prompt.

Let $\mathcal{Q} = \{q_1, q_2, \dots, q_n\}$ be a set of queries. A \emph{routing policy} $\pi : \mathcal{Q} \to \mathcal{R}$ assigns each query to an actor; we write $a_i = \pi(q_i)$ for the actor assigned to query $q_i$.
We denote by $\mathcal{Q}_k = \{q_i \in \mathcal{Q} \mid a_i = r_k\}$ the set of queries assigned to actor $r_k$.

Because actors may differ in architecture and parameterisation,
we do not compare them in weight space.
Instead, we evaluate each actor $r_k$ on an evaluation set
$\mathcal{E} = \{e_1, \dots, e_L\}$, producing a
\emph{behavioural vector},
\begin{equation}
    \mathbf{b}_k =
    \bigl(s(r_k, e_1),\; s(r_k, e_2),\; \dots,\; s(r_k, e_L)\bigr)
    \in \mathbb{R}^{L} \nonumber
    \label{eq:behavioural_vector}
\end{equation}
where $s(r_k, e_l) \in [0, 1]$ is the score of actor $r_k$ on
prompt $e_l$.
Stacking these vectors gives a \emph{behavioural matrix}
$B \in \mathbb{R}^{N \times L}$, from which pairwise inter-actor
distances are computed via cosine similarity (Section~\ref{sec:interagent_distance}).
Differences in parameter space do not necessarily correspond to
differences in behavioural space~\citep{stanley2019designing};
measuring diversity from $B$ rather than from model weights is
therefore both more principled and more general, extending naturally to agent societies that share a model backbone but differ in system prompt or tool access. 
When a separate evaluation set is not available, $\mathcal{E}$
may be set equal to $\mathcal{Q}$.

Given a society $\mathcal{R}$ and a routing policy $\pi$, we
study two properties that are orthogonal to task accuracy.
\textbf{Society diversity}~$\mathrm{HSE}(\mathcal{R})$
(Section~\ref{sec:hse}) quantifies how behaviourally differentiated the
actors in $\mathcal{R}$ are, as measured over the behavioural
matrix $B$. This property is defined over the society alone and
is independent of any routing policy. If $\mathrm{HSE}(\mathcal{R})
\approx 0$, all actors respond similarly and routing is vacuous
regardless of the policy used. \textbf{Routing robustness} $\rho(\pi, \mathcal{Q})$
(Section~\ref{sec:robustness}) quantifies how stably $\pi$ assigns semantically-equivalent query variants to the same actor.
This property is defined over the policy and depends on
$\mathcal{R}$ only through the routing decisions $\pi$ induces. If $\rho(\pi, \mathcal{Q}) \approx 0$, actors receive inconsistent query distributions, which undermines the
conditions under which specialisation can emerge.

Together, these two properties characterise when routing is
\emph{meaningful}: diversity ensures there is something
non-trivial to route to, and robustness ensures that routing decisions satisfy the necessary condition for specialisation.

\section{Society Diversity}
\label{sec:hse}

Given a society $\mathcal{R}$ and its behavioural matrix $B$
(Section~\ref{sec:preliminaries}), we want a single scalar capturing how behaviourally differentiated the actors are.
A useful measure should reflect both the \emph{number} of
distinct behavioural groups and the \emph{degree} to which
those groups differ.  Diversity is ubiquitous in natural systems~\citep{kellert1996value},
and ecologists have demonstrated the role of functional diversity in ecosystem survival~\citep{cadotte2011beyond}.
The challenge of identifying clusters of
elements distributed in a continuous multidimensional space is exactly the problem faced by biologists constructing taxonomic systems, where each dimension represents a morphological trait distinguishing one organism from another.
We adopt the same hierarchical view of entropy, replacing
biological traits with the components of~$\mathbf{b}_k$.

Specifically, we adapt Hierarchic Social Entropy (HSE;~\citealt{balch2000hierarchic}), an information-theoretic measure originally proposed for robot swarms, to LM agent societies. We do not claim diversity is always desirable: for some tasks, homogeneous teams may outperform diverse ones. However, a quantitative metric enables controlled comparisons between societies and informs the design of 
diverse teams.

\begin{comment}
\paragraph{Relation to robotics}
Hierarchic social entropy is an information theoretic measure of robot group diversity introduced by \cite{balch2000hierarchic}. Heterogeneous systems have been growing focus of robotics research and previous work has proposed quantification of robot society heteregeneity \cite{mckee2022quantifying} \cite{wang2019influence} \cite{hu2022policy} \cite{liu2021towards}, but these works focused on discrete spaces. Add paper citations about continuous space diversity metrics. Discuss the System Neural Diversity (SND) paper \cite{bettini2025system}. Explain how they focus on policies. 

Since our models are all different, we cannot compare them at parameter-level, therefore we compare them at behavior-level, evaluating their behavior on a set of prompts. Additionally, differences in parameter space do not necessarily map to differences in behavioral space \cite{stanley2019designing}, hence we measure diversity from model outputs. 
\end{comment}

\subsection{Simple Social Entropy}
\label{sec:simple_entropy}

Simple social entropy applies Shannon's information entropy \cite{shannon1948mathematical} to a multi-agent
society to quantify behavioural diversity based on how actors
are distributed among behavioural subsets.

Let $\mathcal{C} = \{c_1, c_2, \dots, c_M\}$ be a partition of
$\mathcal{R}$ into $M$ homogeneous subsets, and let $p_i$
denote the proportion of actors belonging to subset $c_i$,
with $\sum_{i=1}^{M} p_i = 1$.
The \emph{simple social entropy} of society $\mathcal{R}$
under partition $\mathcal{C}$ is:
\begin{equation}
    H(\mathcal{R}) = -\sum_{i=1}^{M} p_i \log_2(p_i)
    \label{eq:simple_entropy}
\end{equation}

\begin{figure}[t]
\centering
\resizebox{\linewidth}{!}{%
\begin{tikzpicture}[
  circ/.style  = {circle, fill=black, minimum size=8pt, inner sep=0pt},
  starsh/.style= {star, star points=5, fill=black, minimum size=11pt,
                  inner sep=0pt, star point ratio=2.25},
  trish/.style = {regular polygon, regular polygon sides=3, fill=black,
                  minimum size=9pt,  inner sep=0pt, rotate=180},
  sqsh/.style  = {rectangle, fill=black, minimum size=8pt, inner sep=0pt},
]

%--- Box 1 : 12 circles  (H = 0.00) ---
\draw[thick] (0,0) rectangle (2,2);
\foreach \y in {0.28,0.76,1.24,1.72}{
  \foreach \x in {0.38,1.00,1.62}{
    \node[circ] at (\x,\y) {};}}
\node[font=\Large,below] at (1.00, 0) {$0.00$};

%--- Box 2 : 11 circles + 1 star  (H = 0.41) ---
\draw[thick] (2.25,0) rectangle (4.25,2);
% row 0
\node[circ]  at (2.63,0.28){}; \node[circ]  at (3.25,0.28){}; \node[circ]  at (3.87,0.28){};
% row 1
\node[circ]  at (2.63,0.76){}; \node[circ]  at (3.25,0.76){}; \node[circ]  at (3.87,0.76){};
% row 2 — star replaces middle circle
\node[circ]  at (2.63,1.24){}; \node[starsh] at (3.25,1.24){}; \node[circ]  at (3.87,1.24){};
% row 3
\node[circ]  at (2.63,1.72){}; \node[circ]  at (3.25,1.72){}; \node[circ]  at (3.87,1.72){};
\node[font=\Large,below] at (3.25, 0) {$0.41$};

%--- Box 3 : 9 circles + 3 triangles  (H = 0.82) ---
\draw[thick] (4.50,0) rectangle (6.50,2);
% row 0: C C C
\node[circ]  at (4.88,0.28){}; \node[circ]  at (5.50,0.28){}; \node[circ]  at (6.12,0.28){};
% row 1: T C C
\node[trish] at (4.88,0.76){}; \node[circ]  at (5.50,0.76){}; \node[circ]  at (6.12,0.76){};
% row 2: C C T
\node[circ]  at (4.88,1.24){}; \node[circ]  at (5.50,1.24){}; \node[trish] at (6.12,1.24){};
% row 3: C T C
\node[circ]  at (4.88,1.72){}; \node[trish] at (5.50,1.72){}; \node[circ]  at (6.12,1.72){};
\node[font=\Large,below] at (5.50, 0) {$0.82$};

%--- Box 4 : 6 circles + 6 stars  (H = 1.00) ---
\draw[thick] (6.75,0) rectangle (8.75,2);
% row 0: S C S
\node[starsh] at (7.13,0.28){}; \node[circ]   at (7.75,0.28){}; \node[starsh] at (8.37,0.28){};
% row 1: C S C
\node[circ]   at (7.13,0.76){}; \node[starsh] at (7.75,0.76){}; \node[circ]   at (8.37,0.76){};
% row 2: S C S
\node[starsh] at (7.13,1.24){}; \node[circ]   at (7.75,1.24){}; \node[starsh] at (8.37,1.24){};
% row 3: C S C
\node[circ]   at (7.13,1.72){}; \node[starsh] at (7.75,1.72){}; \node[circ]   at (8.37,1.72){};
\node[font=\Large,below] at (7.75, 0) {$1.00$};

%--- Box 5 : 4 circles + 4 stars + 4 triangles  (H = 1.59) ---
\draw[thick] (9.00,0) rectangle (11.00,2);
% row 0: C S T
\node[circ]   at (9.38,0.28){};  \node[starsh] at (10.00,0.28){}; \node[trish]  at (10.62,0.28){};
% row 1: T C S
\node[trish]  at (9.38,0.76){};  \node[circ]   at (10.00,0.76){}; \node[starsh] at (10.62,0.76){};
% row 2: S T C
\node[starsh] at (9.38,1.24){};  \node[trish]  at (10.00,1.24){}; \node[circ]   at (10.62,1.24){};
% row 3: C S T
\node[circ]   at (9.38,1.72){};  \node[starsh] at (10.00,1.72){}; \node[trish]  at (10.62,1.72){};
\node[font=\Large,below] at (10.00, 0) {$1.59$};

%--- Box 6 : 3C + 3S + 3T + 3 squares  (H = 2.00) ---
\draw[thick] (11.25,0) rectangle (13.25,2);
% row 0: C S T
\node[circ]   at (11.63,0.28){}; \node[starsh] at (12.25,0.28){}; \node[trish]  at (12.87,0.28){};
% row 1: Q C S
\node[sqsh]   at (11.63,0.76){}; \node[circ]   at (12.25,0.76){}; \node[starsh] at (12.87,0.76){};
% row 2: T Q C
\node[trish]  at (11.63,1.24){}; \node[sqsh]   at (12.25,1.24){}; \node[circ]   at (12.87,1.24){};
% row 3: S T Q
\node[starsh] at (11.63,1.72){}; \node[trish]  at (12.25,1.72){}; \node[sqsh]   at (12.87,1.72){};
\node[font=\Large, below] at (12.25, 0) {$2.00$};

\end{tikzpicture}%
}
\vspace{-1.8em}
\caption{Simple social entropy computed on societies with different cluster structures. Entropy increases from left to right as agents are distributed  across an increasing number of equal-sized behavioural groups.}
\label{fig:simple_entropy_example}
\end{figure}
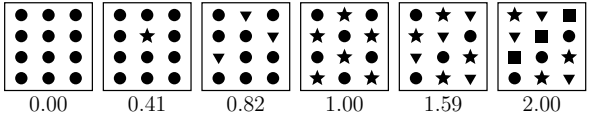

This measure satisfies two basic properties:
(i)~\emph{homogeneity}: $H(\mathcal{R}) = 0$ if and only if
all actors fall into a single subset;
(ii)~\emph{maximum diversity}: $H(\mathcal{R})$ is maximised
when actors are distributed equally across all $M$ subsets
($p_i = 1/M$ for all $i$).
Figure~\ref{fig:simple_entropy_example} illustrates simple
social entropy on societies with varying cluster structure.

\begin{comment}
Simple social entropy applies Shannon's information entropy to multi-agent groups to quantify behavioral diversity based on how agents are distributed among behavioral subsets.

\begin{itemize}
\item Let $\mathcal{R}$ represent a society of $N$ agents, defined as $\mathcal{R}=\{r_{1},r_{2},\dots,r_{N}\}$.

\item Let $\mathcal{C}$ represent a classification partition of $\mathcal{R}$ into $M$ homogeneous subsets, defined as $\mathcal{C}=\{c_{1},c_{2},\dots,c_{M}\}$.

\item Let $p_{i}$ represent the proportion of agents belonging to the $i$-the subset $c_{i}$, where $\sum_{i=1}^{M} p_{i} = 1$.
\end{itemize}

Simple social entropy $H(\mathcal{R})$ of a society is defined as:

$$H(\mathcal{R}) = -\sum_{i=1}^{M} p_{i} \log_{2}(p_{i})$$

The simple social entropy has the following properties: 
\begin{itemize} 
\item Homogeneity: $H(\mathcal{R}) = 0$ if and only if all agents fall into a single subset, representing the minimum possible diversity.

\item Maximum Diversity: For a given number of subsets $M$, $H(\mathcal{R})$ is maximized when agents are distributed equally across all subsets ($p_{i} = \frac{1}{M}$).
\end{itemize}

Figure \ref{fig:simple_entropy_example} illustrates simple social entropy on more of less homogeneous abstract societies. 
\end{comment}

\subsection{Inter-Agent Distance}
\label{sec:interagent_distance}

To partition $\mathcal{R}$ into behavioural subsets, we require
a pairwise distance over actors.
Given the behavioural vectors $\mathbf{b}_k$ 
defined in Section~\ref{sec:preliminaries}, we measure the distance
between actors $r_j$ and $r_k$ as one minus their cosine
similarity:
\begin{equation}
    d(r_j, r_k) = 1 -
    \frac{\mathbf{b}_j \cdot \mathbf{b}_k}
         {\|\mathbf{b}_j\|\,\|\mathbf{b}_k\|}
    \label{eq:distance}
\end{equation}
We prefer cosine similarity over the Euclidean distance used
in the original HSE formulation~\citep{balch2000hierarchic}
because behavioural vectors may differ substantially in
magnitude when agents vary in overall accuracy, and we wish
to capture \emph{profile} similarity rather than absolute
performance level (Appendix \ref{app:linkage}).
We prefer it over the Wasserstein distance used in
SND~\citep{bettini2025system} because that metric is defined
over full action distributions, which are not directly
available for language model outputs.

\subsection{Hierarchic Social Entropy}
\label{sec:hierarchical:hse}

\begin{figure}[t]
\centering
\resizebox{0.7\linewidth}{!}{%
\begin{tikzpicture}[scale=.98,
  circ/.style  = {circle, fill=black, minimum size=9pt, inner sep=0pt},
  starsh/.style= {star, star points=5, fill=black, minimum size=11pt,
                  inner sep=0pt, star point ratio=2.25},
]

%--- Box 1: Close Outlier ---
\draw[thick] (0,0) rectangle (3,3);

% cluster — 3 circles, spread out
\node[circ] at (0.55, 2.42) {};
\node[circ] at (1.35, 2.22) {};
\node[circ] at (0.65, 1.62) {};

% centroid marker (small cross)
\draw[gray, line width=0.5pt]
  (0.85, 2.09) +(-0.08,0) -- +(0.08,0)
  (0.85, 2.09) +(0,-0.08) -- +(0,0.08);

% close outlier — star
\node[starsh] at (2.10, 1.15) {};

% distance line
\draw[dashed, gray, line width=0.6pt] (0.85, 2.09) -- (2.10, 1.15)
  node[midway, right=2pt, font=\small\itshape] {$d_1$};

\node[font=\small, below] at (1.5, 0) {Close Outlier};

%--- Box 2: Far Outlier ---
\draw[thick] (3.5,0) rectangle (6.5,3);

% cluster — same relative positions, offset +3.5
\node[circ] at (4.05, 2.42) {};
\node[circ] at (4.85, 2.22) {};
\node[circ] at (4.15, 1.62) {};

% centroid marker
\draw[gray, line width=0.5pt]
  (4.35, 2.09) +(-0.08,0) -- +(0.08,0)
  (4.35, 2.09) +(0,-0.08) -- +(0,0.08);

% far outlier — star
\node[starsh] at (6.15, 0.40) {};

% distance line
\draw[dashed, gray, line width=0.6pt] (4.35, 2.09) -- (6.15, 0.40)
  node[midway, right=2pt, font=\small\itshape] {$d_2$};

\node[font=\small, below] at (5.0, 0) {Far Outlier};

\end{tikzpicture}%
}
\vspace{-.5em}
\caption{Simple social entropy assigns the same value ($H = 0.811$) to both societies, since it depends only on group proportions, not on inter-group distance. The outlier agent (star) is much farther from the cluster in the right panel ($d_1 \ll d_2$), yet~$H$ is blind to this difference. Hierarchic Social Entropy resolves this by integrating entropy across all taxonomic levels.}
\label{fig:simple_entropy_outlier}
\end{figure}
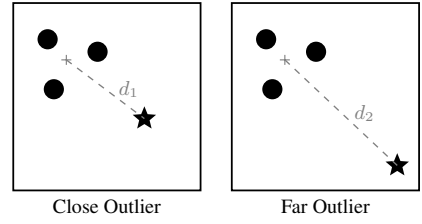

Simple social entropy (Equation~\ref{eq:simple_entropy}) captures two aspects of diversity: the number of behavioural groups and the distribution of actors across them.
It does not, however, capture the \emph{degree} of difference
between groups. Figure~\ref{fig:simple_entropy_outlier} illustrates the consequence: two societies each containing three identical actors and one outlier receive the same score $H = 0.811$, regardless of whether the outlier is close to or far from the main cluster ($d_1 \ll d_2$, yet $H$ is identical). This limitation matters for routing. Suppose a router assigns queries to an outlier actor that barely differs from the rest of the society. Simple entropy would record non-zero diversity, yet the routing decision is nearly vacuous: the outlier provides almost no complementary capability.
A useful diversity measure should be sensitive to the
\emph{degree} of behavioural separation between groups,
not just to whether separation exists.

To obtain a single society-level score, we aggregate 
pairwise inter-actor distances (defined in Section~\ref{sec:interagent_distance}). \citet{bettini2025system} aggregate by taking the mean over all
pairwise distances.
This is computationally simple but conflates societies with
very different cluster structures: a society with two tight,
well-separated clusters and a society with uniformly scattered
actors can yield the same mean distance, yet the former is far
more useful for routing, since actors fall into stable,
distinguishable groups. \citet{balch2000hierarchic} instead cluster actors at every possible distance threshold and measure entropy at each scale. We adopt this approach for the same reason: it is sensitive to
the hierarchic \emph{structure} of the behavioural space, not just its
average extent.

Specifically, we use hierarchical clustering with single linkage
(nearest-point algorithm), which merges two clusters as soon as their closest members fall within threshold~$h$.
This is appropriate for our setting, where even a single
behaviourally distinct actor justifies a separate cluster.\footnote{
The original HSE formulation uses Euclidean distance and centroid linkage rather than
cosine distance and single linkage; we compare both in Appendix~\ref{app:linkage}.}
As $h$ varies, the partition $\mathcal{C}$ changes, so simple
social entropy becomes a function of both the society and $h$,
written $H(\mathcal{R}, h)$. We call $h$ the \emph{taxonomic level}, borrowing the term from
\citet{balch2000hierarchic}, who in turn draws on biological
taxonomy: just as a biologist can classify organisms at the
species level or the genus level depending on how finely they
wish to distinguish, $h$ controls the resolution at which we
observe the society. At $h = 0$, actors merge only if their distance is exactly
zero; behaviourally distinct actors each form a singleton
cluster, giving $H(\mathcal{R}, 0) = \log_2 M$ where
$M \leq N$ is the number of distinct behavioural profiles.
As $h$ increases, similar actors merge and the number of
clusters falls, until at $h \to \infty$ all actors collapse
into one group with $H(\mathcal{R}, h) = 0$.

\emph{Hierarchic Social Entropy} integrates simple entropy
across all taxonomic levels:
\begin{equation}
    \mathrm{HSE}(\mathcal{R}) = \int_{0}^{\infty} H(\mathcal{R}, h)\, dh
    \label{eq:hse}
\end{equation}
We define $d(r_j, r_k) = 1$ (maximum distance) when either 
$\|\mathbf{b}_j\| = 0$ or $\|\mathbf{b}_k\| = 0$, i.e., when 
an actor fails every evaluation prompt.
HSE is a continuous ratio measure with an absolute zero (when all
actors are identical), enabling statements of the form ``Society A is
twice as diverse as Society B'' and total orderings across societies
of different sizes when normalised by the maximum achievable HSE for
that society size.

\subsection{Validating HSE}

\input{table/hse_compact1}

\label{sec:simple_examples}

We construct minimal societies that isolate specific properties
of HSE, confirming that the metric behaves as expected before
applying it at scale (see Section~\ref{sec:results}).
Table~\ref{tab:hse-sanity} verifies that when all actors produce identical outputs (a), the society is
maximally homogeneous: HSE is zero and routing is vacuous,
regardless of how many actors are present.
When actors are fully orthogonal, each succeeding on exactly
the prompts the others fail (b), HSE is maximised and
normalised HSE equals 1.0.
Crucially, the normalised value is invariant to society size,
enabling fair comparisons across societies of different scales.

Table~\ref{tab:hse-duplicates} verifies that adding redundant
information does not artificially inflate HSE.
Starting from a base society of three behaviourally distinct
actors (case~a), we extend it in two ways: adding a fourth actor
whose behavioural vector is identical to actor~C (case~b), and
adding a fourth evaluation prompt that is a copy of~p3 (case~c).
In both cases HSE decreases slightly rather than increasing,
confirming that the metric is not fooled by redundancy.
This property matters in practice: a routing system should not
appear more diverse simply because actors are duplicated or
because the evaluation set contains repeated queries.
Finally, Figure~\ref{fig:hse_incremental} empirically confirms monotonicity: HSE increases as behaviourally distinct actors are added.

\input{table/hse_compact2}

\section{Robustness of Routing Policy}
\label{sec:robustness}

Society diversity (Section~\ref{sec:hse}) characterises
whether a society contains actors with sufficiently
differentiated capabilities to make routing non-trivial.
But diversity alone is not sufficient: as the specialisation
literature makes clear (Section~\ref{sec:related_work}),
actors can only develop stable roles if they reliably receive
the same type of queries over time.

We now introduce a metric that captures this second
structural property directly.

The central intuition is that stable query assignment is a
necessary  condition for specialisation to be
possible: a router that assigns query $q_i$ to actor $r_k$
but assigns a surface-form variant to actor $r_j \neq r_k$
violates this condition, regardless of what learning
dynamics operate downstream. We do not claim that stable
routing produces specialisation --- whether actors develop
specialised competence depends on factors outside the scope
of this work --- but unstable routing makes it impossible
by construction. We therefore measure routing robustness
as the degree to which surface-form variants of a query
are assigned to the same actor.

\subsection{Perturbation Taxonomy}
\label{sec:perturbations}

The robustness metric $\rho$ is defined relative to a
perturbation regime: the same router may be stable under
surface-level noise yet unstable under semantic
reformulation.
To support both aggregate and fine-grained analysis, we
define five perturbation classes of increasing semantic
distance from the original query, summarised in
Table~\ref{tab:perturbations}.

\begin{table}[t]
\centering
\small
\begin{tabular}{@{}lll@{}}
\toprule
\textbf{Level} & \textbf{Class} & \textbf{Example} \\
\midrule
1 & Character-level  & Typos, spacing, punctuation \\
2 & Word-level       & Synonym substitution \\
3 & Syntax-level     & Passive voice, clause reordering \\
4 & Paraphrase       & Full rewrite, same meaning \\
5 & Rambling         & Added irrelevant context \\
\bottomrule
\end{tabular}
\vspace{-.5em}
\caption{Perturbation taxonomy, ordered by increasing
semantic distance from the original query.}
\label{tab:perturbations}
\end{table}

Perturbations at levels 1--3 preserve surface proximity
to the original query; levels 4--5 test whether the router
is sensitive to meaning-preserving reformulations that
substantially alter wording or add noise.
Full methodology and prompt templates used to generate each class of perturbations are provided
in Appendix~\ref{app:perturbation_templates}.
By computing $\rho$ separately for each perturbation class,
one can characterise \emph{where} a routing policy breaks
down: a router that degrades only at level 4--5 is more
robust in practice than one that degrades at level 2,
even if their aggregate $\rho$ scores are similar.

\subsection{Routing Robustness Metric}
\label{sec:robustness_metric} 

Given a routing policy $\pi$ and a query $q_i \in \mathcal{Q}$,
recall that $a_i = \pi(q_i)$ denotes the actor assigned to
the original query (Section~\ref{sec:preliminaries}).
We generate $p$ perturbations $q_i^1, q_i^2, \ldots, q_i^p$
of $q_i$ (see Section~\ref{sec:perturbations} for the
perturbation taxonomy).
The \textbf{per-query robustness} is the fraction of
perturbations routed to the same actor as the original:
\begin{equation}
    \rho_i = \frac{1}{p} \sum_{j=1}^{p}
          \mathbf{1}\!\left[\pi(q_i^j) = a_i\right]
    \label{eq:per_query_robustness}
\end{equation}
where $\mathbf{1}[\cdot]$ is the indicator function.
$\rho_i \in [0,1]$, with $\rho_i = 1$ if all perturbations
are routed to the same actor, and $\rho_i = 0$ if none are.

\textbf{Dataset-level robustness} is obtained by averaging over all
$n = |\mathcal{Q}|$ queries:
\begin{equation}
    \rho(\pi, \mathcal{Q}) = \frac{1}{n} \sum_{i=1}^{n} \rho_i
    \label{eq:dataset_robustness}
\end{equation}
This metric rests on the assumption that meaning-preserving 
perturbations do not change which actor is most accurate for 
a given query: if $q_i$ and $q_i^j$ are semantically 
equivalent, the optimal assignment should be the same for 
both. Under this assumption, any change in routing decision 
reflects instability in the policy rather than a legitimate 
response to a change in query difficulty or actor suitability.

Two boundary cases anchor the metric: a \emph{fixed router} that always assigns the same actor gives \mbox{$\rho_i = 1$} for all $i$ and  $\rho = 1$, while a \emph{uniform random router} gives $\mathbb{E}[\rho] = 1/N$. Beyond the dataset-level aggregate, we also compute a per-actor  score over the query
set $\mathcal{Q}_k$ assigned to actor $r_k$:
\begin{equation}
    \rho^{(k)}(\pi, \mathcal{Q}) =
    \frac{1}{|\mathcal{Q}_k|}
    \sum_{q_i \in \mathcal{Q}_k} \rho_i
    \label{eq:per_actor_robustness}
\end{equation}
If no queries are assigned to actor $r_k$ under policy~$\pi$  (i.e., $|\mathcal{Q}_k| = 0$), we define $\rho^{(k)}(\pi, \mathcal{Q}) = 0$.
A low $\rho^{(k)}$ indicates that queries nominally
assigned to actor $r_k$ are unstably routed under
perturbation, violating the necessary condition for
specialisation to emerge for that actor's role.

\section{Experimental Setup}
\label{sec:expsetup}

\paragraph{Datasets}
We evaluate on two routing benchmarks.
\textbf{EmbedLLM}~\citep{zhuang2025embedllm} provides performance scores across 112 models on 10 reasoning and knowledge benchmarks
(MMLU, GSM8K, GPQA, ASDiv, and others), comprising 36,054 questions in total.  \textbf{RouterBench}~\citep{hu2024routerbench} is a
complementary routing benchmark with performance scores across 11~ models on 36,497 queries drawn from 8~datasets (see
Appendix~\ref{sec:router:benchmarks} for details and train, validation, and test splits). 

\paragraph{Societies}\hspace{-.2cm}The benchmarks described above constitute the \emph{default societies}. On EmbedLLM, we compare the default society  (HSE=1.62) against three synthetic societies in which  models are replaced by synthetic experts with binary performance indicators encoding domain-specific expertise:

\begin{itemize}[itemsep=0pt, labelindent=0pt, leftmargin=*]

    \item \textbf{RD} (Reasoning Depth): five non-overlapping experts covering five tiers of cognitive complexity, ranging from direct lookup to complex multi-step reasoning (HSE\,=\,2.32).
    \item \textbf{SA} (Subject Area): 15 non-overlapping experts each specialising in a distinct academic or practical domain, such as advanced mathematics, computer science, medicine, and commonsense reasoning (HSE\,=\,3.91).
    \item \textbf{RD+SA}: the combined pool of 20 RD and SA experts with partially overlapping expertise (HSE\,=\,3.19).
\end{itemize}

Expert categorisation heuristics are detailed in
Appendix~\ref{sec:synthetic}. Oracle accuracy over each synthetic society is 1.0 by construction, since each query falls within exactly one expert's domain. We define a \emph{max-HSE subset} for each benchmark: the subset of the full society that maximises normalised HSE at a fixed society size, selected greedily; this subset serves as a high-diversity reference within real-world model pools. On
RouterBench we compare the full society against its max-HSE subset.

\paragraph{Routing Policies}
We evaluate five routing policies: \textbf{random} assigns each query to a uniformly sampled actor; \textbf{prompted} prompts an LLM with
natural-language descriptions of actor roles and few-shot examples to select the most
appropriate one (see Appendix \ref{app:prompted_router}); \textbf{KNN-$k$} ($k \in \{1, 3, 10\}$) retrieves the
$k$ nearest neighbours of the query in embedding space and assigns the
actor that performed best on those neighbours (Appendix \ref{app:knn_router}).

\paragraph{Perturbations}
Routing robustness (Section~\ref{sec:robustness_metric}) is measured
over the five-level perturbation taxonomy of
Section~\ref{sec:perturbations}. For each query we generate one
perturbation per level, yielding $p = 5$ variants (Appendix \ref{app:perturbation_templates}); we report
dataset-level robustness $\rho(\pi, \mathcal{Q})$
(Equation~\ref{eq:dataset_robustness}) and per-level breakdowns.

\section{Results and Analysis}
\label{sec:results}

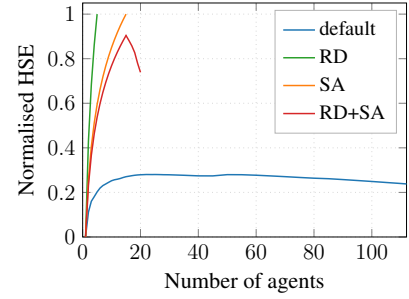
\begin{figure}[t]
\centering
\begin{tikzpicture}[scale=.7]
\begin{axis}[
  width=\linewidth,
  height=6.0cm,
  xmin=0, xmax=112,
  ymin=0.0, ymax=1.05,
  xtick={0,20,40,60,80,100},
  ytick={0,0.2,0.4,0.6,0.8,1.0},
  xticklabel style={font=\large},
  yticklabel style={font=\large,
    /pgf/number format/fixed,
    /pgf/number format/precision=1},
  xlabel={Number of agents},
  ylabel={Normalised HSE},
  xlabel style={font=\large},
  ylabel style={font=\large},
  grid=major,
  grid style={dotted, gray!40},
  legend pos=north east,
  legend style={font=\normalsize, draw=gray!60, fill=white,
    row sep=2pt, inner sep=4pt},
  legend cell align=left,
  every axis plot/.append style={thick, mark size=0pt},
]

%% default society (blue)
\addplot[mpblue]
  coordinates {
    (1,0.0000)(2,0.1117)(3,0.1602)(4,0.1808)(5,0.2021)
    (6,0.2194)(7,0.2307)(8,0.2380)(9,0.2454)(10,0.2524)
    (11,0.2567)(12,0.2592)(13,0.2625)(14,0.2669)(15,0.2705)
    (16,0.2727)(17,0.2750)(18,0.2769)(19,0.2783)(20,0.2792)
    (21,0.2797)(22,0.2803)(23,0.2802)(24,0.2801)(25,0.2803)
    (26,0.2801)(27,0.2798)(28,0.2794)(29,0.2793)(30,0.2791)
    (35,0.2769)(40,0.2743)(45,0.2741)(50,0.2797)(55,0.2794)
    (60,0.2771)(65,0.2740)(70,0.2707)(75,0.2671)(80,0.2638)
    (85,0.2615)(90,0.2576)(95,0.2534)(100,0.2491)(105,0.2445)
    (110,0.2400)(112,0.2380)
  };
\addlegendentry{default}

%% synthetic_rd (green) — 5 agents only
\addplot[mpgreen]
  coordinates {
    (1,0.0000)(2,0.4307)(3,0.6826)(4,0.8614)(5,1.0000)
  };
\addlegendentry{RD}

%% synthetic_sa (orange) — 15 agents only
\addplot[mporange]
  coordinates {
    (1,0.0000)(2,0.2560)(3,0.4057)(4,0.5119)(5,0.5943)
    (6,0.6616)(7,0.7186)(8,0.7679)(9,0.8114)(10,0.8503)
    (11,0.8855)(12,0.9176)(13,0.9472)(14,0.9745)(15,1.0000)
  };
\addlegendentry{SA}

%% synthetic_sa_rd (red) — 20 agents
\addplot[mpred]
  coordinates {
    (1,0.0000)(2,0.2314)(3,0.3667)(4,0.4628)(5,0.5372)
    (6,0.5981)(7,0.6496)(8,0.6941)(9,0.7335)(10,0.7686)
    (11,0.8004)(12,0.8295)(13,0.8562)(14,0.8809)(15,0.9040)
    (16,0.8817)(17,0.8578)(18,0.8274)(19,0.7793)(20,0.7389)
  };
\addlegendentry{RD+SA}

\end{axis}
\end{tikzpicture}
 \vspace{-8pt}
\caption{Normalised HSE as a function of society size for the default
  (real-world) and synthetic societies on EmbedLLM, using cosine
  distance and single linkage (greedy agent selection).
  All societies exhibit strong diminishing returns; the default society
  plateaus after approximately nine agents.}
\label{fig:hse_incremental}
\vspace{-10pt}
\end{figure}

\paragraph{Specialist societies are substantially more diverse.}
Figure~\ref{fig:hse_incremental} shows normalised HSE as a function of society size on EmbedLLM. At any fixed size, purpose-designed specialist societies achieve higher HSE than the default real-world model pool. The SA society (HSE\,=\,3.91) is more than twice as diverse as the default (HSE\,=\,1.62), and the RD+SA combined pool (HSE\,=\,3.19) similarly outpaces it despite containing many fewer total actors. This indicates that a large number of real-world models occupy similar regions of behavioural space: size alone does not imply diversity.

Across all societies, HSE exhibits strong diminishing returns: the marginal gain from adding an actor falls sharply after the first few. For the default society, the curve plateaus after approximately nine agents. Synthetic specialist societies plateau even faster, at around four to five agents for RD and SA,  because their behavioural profiles are maximally orthogonal by construction. Small, carefully curated societies can recover most of the diversity available in a much larger pool, and collecting additional models offers little incremental benefit. HSE can also be combined with task accuracy as a joint
selection criterion: Appendix~\ref{sec:model:selection}
reports results for subsets selected by maximising HSE
jointly with task accuracy, showing that the joint
criterion leads to better routing  while
preserving diversity.

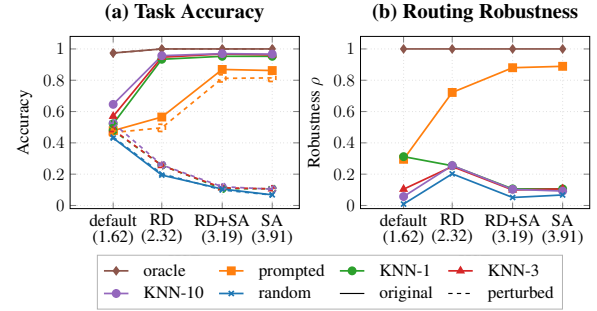
\begin{figure}[t]
\centering
\resizebox{\linewidth}{!}{%
\begin{tikzpicture}
\begin{groupplot}[
  group style={group size=2 by 1, horizontal sep=1.5cm},
  width=7.2cm, height=5.8cm,
  xmin=1.0, xmax=4.3,
  ymin=-0.02, ymax=1.08,
  xtick={1.62, 2.32, 3.19, 3.91},
  xticklabels={
    \shortstack{\large default\\\large(1.62)},
    \shortstack{\large RD\\\large(2.32)},
    \shortstack{\large RD+SA\\\large(3.19)},
    \shortstack{\large SA\\\large(3.91)}
  },
  xticklabel style={align=center, font=\large},
  xlabel={HSE},
  xlabel style={font=\large},
  ytick={0,0.2,0.4,0.6,0.8,1.0},
  yticklabel style={font=\large},
  grid=major,
  grid style={dotted, gray!40},
  every axis plot/.append style={mark size=2.5pt},
]

%── (a) Task Accuracy ────────────────────────────────────────────
\nextgroupplot[
  title={\Large \textbf{(a) Task Accuracy}},
  title style={font=\large},
  ylabel={\large Accuracy},
  ylabel style={font=\large},
  legend to name=bottomlegend,
  legend columns=4,
  legend style={
    font=\large,
    draw=gray!60,
    fill=white,
    row sep=2pt,
    inner sep=4pt,
    column sep=8pt,
  },
  legend cell align=left,
]

% oracle
\addplot[mpbrown, very thick, mark=diamond*]
  coordinates {(1.62,0.974)(2.32,1.000)(3.19,1.000)(3.91,1.000)};
\addlegendentry{oracle}
\addplot[mpbrown, very thick, dashed,
  mark=diamond, every mark/.append style={fill=white}, forget plot]
  coordinates {(1.62,0.974)(2.32,1.000)(3.19,1.000)(3.91,1.000)};

% prompted
\addplot[mporange, very thick, mark=square*]
  coordinates {(1.62,0.478)(2.32,0.565)(3.19,0.869)(3.91,0.862)};
\addlegendentry{prompted}
\addplot[mporange, very thick, dashed,
  mark=square, every mark/.append style={fill=white}, forget plot]
  coordinates {(1.62,0.466)(2.32,0.496)(3.19,0.812)(3.91,0.815)};

% KNN-1
\addplot[mpgreen, very thick, mark=*]
  coordinates {(1.62,0.522)(2.32,0.934)(3.19,0.953)(3.91,0.953)};
\addlegendentry{KNN-1}
\addplot[mpgreen, very thick, dashed,
  mark=o, every mark/.append style={fill=white}, forget plot]
  coordinates {(1.62,0.479)(2.32,0.257)(3.19,0.107)(3.91,0.107)};

% KNN-3
\addplot[mpred, very thick, mark=triangle*]
  coordinates {(1.62,0.570)(2.32,0.947)(3.19,0.968)(3.91,0.962)};
\addlegendentry{KNN-3}
\addplot[mpred, very thick, dashed,
  mark=triangle, every mark/.append style={fill=white}, forget plot]
  coordinates {(1.62,0.484)(2.32,0.255)(3.19,0.113)(3.91,0.106)};

% KNN-10
\addplot[mppurple, very thick, mark=oplus*]
  coordinates {(1.62,0.646)(2.32,0.957)(3.19,0.970)(3.91,0.967)};
\addlegendentry{KNN-10}
\addplot[mppurple, very thick, dashed,
  mark=oplus, every mark/.append style={fill=white}, forget plot]
  coordinates {(1.62,0.526)(2.32,0.259)(3.19,0.120)(3.91,0.104)};

% random
\addplot[mpblue, very thick, mark=x]
  coordinates {(1.62,0.431)(2.32,0.194)(3.19,0.105)(3.91,0.068)};
\addlegendentry{random}
\addplot[mpblue, very thick, dashed, mark=x, forget plot]
  coordinates {(1.62,0.437)(2.32,0.202)(3.19,0.099)(3.91,0.067)};

\addlegendimage{black, thick}
\addlegendentry{original}
\addlegendimage{black, thick, dashed}
\addlegendentry{perturbed}

%── (b) Routing Robustness ───────────────────────────────────────
\nextgroupplot[
  title={\Large \textbf{(b) Routing Robustness}},
  title style={font=\large},
  ylabel={\large Robustness $\rho$},
  ylabel style={font=\large},
  xlabel={\large HSE},
  xlabel style={font=\large},
]

\addplot[mpbrown, very thick, mark=diamond*]
  coordinates {(1.62,1.000)(2.32,1.000)(3.19,1.000)(3.91,1.000)};
\addplot[mporange, very thick, mark=square*]
  coordinates {(1.62,0.295)(2.32,0.722)(3.19,0.880)(3.91,0.889)};
\addplot[mpgreen, very thick, mark=*]
  coordinates {(1.62,0.312)(2.32,0.254)(3.19,0.106)(3.91,0.106)};
\addplot[mpred, very thick, mark=triangle*]
  coordinates {(1.62,0.104)(2.32,0.249)(3.19,0.099)(3.91,0.106)};
\addplot[mppurple, very thick, mark=oplus*]
  coordinates {(1.62,0.057)(2.32,0.255)(3.19,0.103)(3.91,0.092)};
\addplot[mpblue, very thick, mark=x]
  coordinates {(1.62,0.009)(2.32,0.202)(3.19,0.051)(3.91,0.067)};

\end{groupplot}

%% Shared legend below both panels
\node[anchor=north] at
  ($(group c1r1.south east)!0.5!(group c2r1.south west) - (0,1.0cm)$)
  {\pgfplotslegendfromname{bottomlegend}};

\end{tikzpicture}
}
\vspace{-.7cm}

\caption{Task accuracy (a) and  robustness $\rho$ (b) across EmbedLLM societies ordered by HSE. Solid lines show original queries; dashed lines in (a) show perturbations. KNN routers gain accuracy on specialist societies but lose robustness; the prompted router maintains both.}
\label{fig:score_embedllm_synthetic}
\vspace{-10pt}
\end{figure}

\paragraph{KNN routers gain accuracy from specialist societies but collapse in robustness.}
Figure~\ref{fig:score_embedllm_synthetic}(a) shows task accuracy rising sharply and monotonically for KNN routers as HSE increases. However, perturbed-query accuracy collapses at every specialist society, falling to near-random levels. Panel~(b) confirms that robustness $\rho$  remains low across the entire HSE range despite high clean-query accuracy. The sharp behavioural boundaries that make specialist societies easy to route correctly on clean queries make them equally easy to route \emph{incorrectly} when a surface-level reformulation shifts the query embedding across a domain boundary.

The prompted router shows the reverse pattern. Its clean accuracy on SA is lower than KNN-10, but perturbed accuracy remains close, a gap of less than five points. Crucially, robustness rises monotonically with HSE, tracking  society diversity  in a way no KNN router does. Prompted routing  latches onto the semantic intent of the query rather than its embedding position, making it naturally stable under meaning-preserving reformulations.

\begin{figure}[t]
\centering
\resizebox{\linewidth}{!}{%
\begin{tikzpicture}
\begin{groupplot}[
  group style={group size=2 by 1, horizontal sep=2cm},
   width=7.2cm, height=5.8cm,
  xmin=-0.35, xmax=1.35,
  ymin=-0.02, ymax=1.08,
  xtick={0,1},
  ytick={0,0.2,0.4,0.6,0.8,1.0},
  yticklabel style={font=\large},
  grid=major,
  grid style={dotted, gray!40},
  every axis plot/.append style={mark size=3pt, thick},
]
\nextgroupplot[
  title={\Large \textbf{(a) EmbedLLM}},
  title style={font=\large},
  ylabel={Robustness $\rho$},
  ylabel style={font=\large},
  xticklabels={
    \shortstack{\large Default\\\large(0.24)},
    \shortstack{\large Max-HSE\\\large(0.53)}
  },
  xticklabel style={align=center, font=\large},
  legend to name=rholegend,
  legend columns=5,
  legend style={font=\scriptsize, draw=gray!60, fill=white,
    row sep=2pt, inner sep=4pt, column sep=8pt},
  legend cell align=left,
]
\addplot[mpblue,   mark=*]         coordinates {(0,0.009)(1,0.111)};
\addlegendentry{\large random}
\addplot[mporange, mark=square*]   coordinates {(0,0.295)(1,0.613)};
\addlegendentry{\large prompted}
\addplot[mpgreen,  mark=triangle*] coordinates {(0,0.312)(1,0.205)};
\addlegendentry{\large KNN-1}
\addplot[mpred,    mark=diamond*]  coordinates {(0,0.104)(1,0.180)};
\addlegendentry{\large KNN-3}
\addplot[mppurple, mark=oplus*]    coordinates {(0,0.057)(1,0.270)};
\addlegendentry{\large KNN-10}

\nextgroupplot[
  title={\Large \textbf{(b) RouterBench}},
  title style={font=\large},
   ylabel={Robustness $\rho$},
  ylabel style={font=\large},
  xticklabels={
    \shortstack{\large Default \\\large(0.26)},
    \shortstack{\large Max-HSE\\\large(0.44)}
  },
  xticklabel style={align=center, font=\large},
]
\addplot[mpblue,   mark=*]         coordinates {(0,0.091)(1,0.248)};
\addplot[mporange, mark=square*]   coordinates {(0,0.603)(1,0.719)};
\addplot[mpgreen,  mark=triangle*] coordinates {(0,0.307)(1,0.700)};
\addplot[mpred,    mark=diamond*]  coordinates {(0,0.168)(1,0.664)};
\addplot[mppurple, mark=oplus*]    coordinates {(0,0.274)(1,0.729)};

\end{groupplot}

%% Shared legend below both panels
\node[anchor=north] at
  ($(group c1r1.south east)!0.5!(group c2r1.south west) - (0,1.0cm)$)
  {\pgfplotslegendfromname{rholegend}};

\end{tikzpicture}%
}
\vspace{-.7cm}
\caption{Routing robustness $\rho$ on the default society vs.\ the max-HSE subset for EmbedLLM (a) and RouterBench (b). Switching to the max-HSE subset improves the prompted router on both benchmarks, and substantially improves all KNN routers on RouterBench. On EmbedLLM, KNN-1 robustness decreases.}
\vspace{-10pt}
\label{fig:rho_by_router}
\end{figure}
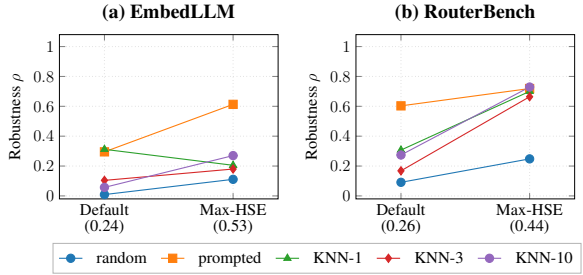

\paragraph{Max-HSE subsets improve prompted robustness, but the effect depends on pool structure.}
Figure~\ref{fig:rho_by_router} compares the full society against the max-HSE subset on EmbedLLM and RouterBench. The prompted router improves consistently, roughly doubling its robustness on EmbedLLM and improving more modestly on RouterBench. The KNN routers tell a more complex story. On RouterBench, switching to the max-HSE subset dramatically rescues KNN robustness across all values of $k$. On EmbedLLM, however, KNN-1 robustness actually decreases while KNN-3 and KNN-10 improve only marginally. This contrast reflects the nature of the max-HSE subset in each case: the RouterBench subset selects from models with smooth, graded behavioural profiles, so higher diversity sharpens actor distinctions without producing hard domain boundaries; the EmbedLLM pool, even at maximum HSE, has less structured behavioural separation, leaving nearest-neighbour routing sensitive to small surface variations in the query.

\begin{figure}[t]
\centering
\resizebox{\linewidth}{!}{%
\begin{tikzpicture}
\begin{groupplot}[
  group style={group size=2 by 1, horizontal sep=2cm},
  width=7.2cm, height=5.8cm,
  xmin=0.5, xmax=5.5,
  ymin=-0.02, ymax=1.08,
  xtick={1,2,3,4,5},
  xticklabels={char, word, syntax, para, rambling},
  xticklabel style={font=\large, rotate=30, anchor=north east},
  ytick={0,0.2,0.4,0.6,0.8,1.0},
  yticklabel style={font=\large},
  grid=major,
  grid style={dotted, gray!40},
  every axis plot/.append style={mark size=2.8pt, thick},
  xlabel={Perturbation level},
  xlabel style={font=\large},
]
\nextgroupplot[
  title={\Large \textbf{(a) Default society}},
  title style={font=\large},
  ylabel={Robustness $\rho$},
  ylabel style={font=\large},
  legend to name=pertlegend,
  legend columns=5,
  legend style={font=\large, draw=gray!60, fill=white,
    row sep=2pt, inner sep=4pt, column sep=8pt},
  legend cell align=left,
]
\addplot[mpblue,   mark=triangle*]
  coordinates {(1,0.093)(2,0.092)(3,0.092)(4,0.088)(5,0.092)};
\addlegendentry{random}
\addplot[mporange, mark=square*]
  coordinates {(1,0.604)(2,0.614)(3,0.617)(4,0.589)(5,0.593)};
\addlegendentry{prompted}
\addplot[mpgreen,  mark=*]
  coordinates {(1,0.298)(2,0.309)(3,0.311)(4,0.304)(5,0.313)};
\addlegendentry{KNN-1}
\addplot[mpred,    mark=diamond*]
  coordinates {(1,0.163)(2,0.167)(3,0.170)(4,0.165)(5,0.173)};
\addlegendentry{KNN-3}
\addplot[mppurple, mark=oplus*]
  coordinates {(1,0.275)(2,0.275)(3,0.283)(4,0.275)(5,0.261)};
\addlegendentry{KNN-10}

\nextgroupplot[
  title={\Large \textbf{(b) Max-HSE subset}},
  title style={font=\large},
   ylabel={Robustness $\rho$},
  ylabel style={font=\large},
]
\addplot[mporange, mark=square*]
  coordinates {(1,0.736)(2,0.744)(3,0.731)(4,0.663)(5,0.720)};
\addplot[mpgreen,  mark=*]
  coordinates {(1,0.697)(2,0.695)(3,0.702)(4,0.711)(5,0.693)};
\addplot[mppurple, mark=oplus*]
  coordinates {(1,0.725)(2,0.717)(3,0.730)(4,0.743)(5,0.730)};
\addplot[mpred,    mark=diamond*]
  coordinates {(1,0.657)(2,0.655)(3,0.670)(4,0.680)(5,0.655)};
\addplot[mpblue,   mark=triangle*]
  coordinates {(1,0.248)(2,0.248)(3,0.246)(4,0.253)(5,0.248)};

\end{groupplot}

%% Shared legend below both panels
\node[anchor=north] at
  ($(group c1r1.south east)!0.5!(group c2r1.south west) - (0,1.2cm)$)
  {\pgfplotslegendfromname{pertlegend}};

\end{tikzpicture}%
}
\vspace{-.7cm}
\caption{Routing robustness $\rho$ per perturbation level on RouterBench for the default, full society (a) and the max-HSE subset (b). All routers are approximately flat across the five levels, with no consistent degradation from character noise to full paraphrase or rambling.}
\label{fig:rho_by_category}
\vspace{-10pt}
\end{figure}
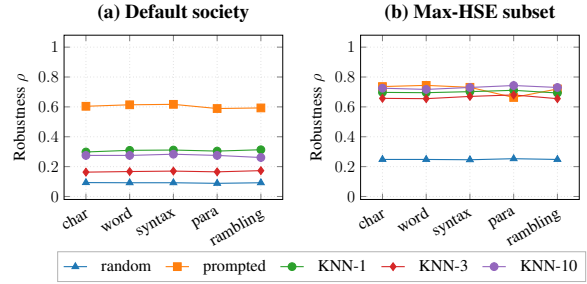

\paragraph{Robustness is roughly flat across perturbation levels.}
Figure~\ref{fig:rho_by_category} shows per-level robustness on RouterBench. Every router produces approximately the same $\rho$ at all five levels: the prompted router ranges from 0.589 (paraphrase) to 0.617 (syntax) on the full society, and from 0.663 (paraphrase) to 0.744 (word) on the max-HSE subset. No router shows a consistent decline as perturbations become semantically deeper. Routing policies are no more sensitive to a full paraphrase than to a character-level typo, implying that routing decisions are not anchored to surface features that higher-level perturbations would selectively disrupt.
The only hint of level-sensitivity is the prompted router's paraphrase dip on the max-HSE subset (0.663 vs.\ $\approx\!0.73$ elsewhere), suggesting that full rewrites may destabilise an LLM-based policy when actor distinctions are sharper.

\section{Conclusion}
We have argued that routing in language-model societies should be evaluated along two structural dimensions beyond task accuracy: behavioural diversity and routing stability under surface-form query variants. For diversity, we adapted Hierarchic Social Entropy to language-model societies, using cosine similarity over behavioural vectors as the inter-actor distance. For robustness, we introduced a perturbation-based metric measuring the fraction of surface-form variants assigned to the same actor.

Applied to EmbedLLM and RouterBench, our analysis  finds that HSE exhibits strong diminishing returns: fewer than ten agents capture most available diversity, providing a practical heuristic for society design. In addition,  robustness and accuracy trade off sharply: KNN routers achieve high accuracy on specialist societies but collapse under perturbation, whereas prompted routers are less accurate but more stable across perturbation types and society configurations. 

Together, HSE and routing robustness ask not only whether a router performs well, but whether routing is meaningful: whether the society is sufficiently differentiated to justify routing decisions, and whether the policy  satisfies the necessary condition for specialisation.

\section{Limitations}

\paragraph{Measurement assumptions}
HSE is computed over a fixed evaluation set $\mathcal{E}$; if this set
is not representative of the actual query distribution, behavioural
vectors may not reflect deployment behaviour and diversity estimates
could be misleading. The max-HSE subset is selected greedily, which
does not guarantee the globally optimal subset at a given size; 
the true optimum requires an exponential search. Our synthetic societies
further use binary performance indicators that produce maximally sharp
behavioural boundaries and oracle accuracy of 1.0 by construction;
real specialist agents will exhibit softer, overlapping profiles,
and the KNN brittleness we observe may be attenuated in practice.

\paragraph{Scope of the robustness metric}
Routing robustness is measured at the assignment level over single-turn
queries and does not capture whether actors actually develop specialised
competence as a result of stable query distributions. Extending the metric to multi-step agentic settings, where
perturbations propagate through state rather than through a single query,
is a natural next step, as is using HSE as a training signal to bootstrap
diverse societies from a homogeneous initialisation.

\bibliography{main}

\newpage

\include{appendix}

\end{document}

%% file: table/hse_compact1.tex
% ── Table 1: sanity checks ──────────────────────────────────────
\begin{table}[t]
\centering
\label{tab:hse-sanity}

\subcaptionbox{Three identical models\newline
               \small\textbf{HSE:}~0.0\quad\textbf{Norm:}~0.0}{%
  \begin{tabular}{@{}lccc@{}}
  \toprule
  \textbf{Prompt} & \textbf{A} & \textbf{B} & \textbf{C} \\
  \midrule
  p1 & 1 & 1 & 1 \\
  p2 & 0 & 0 & 0 \\
  p3 & 0 & 0 & 0 \\
  \bottomrule
  \end{tabular}}%
\hspace*{.3cm}
\subcaptionbox{Three orthogonal models\newline
               \small\textbf{HSE:}~1.585\quad\textbf{Norm:}~1.0}{%
  \begin{tabular}{lccc@{}}
  \toprule
  \textbf{Prompt} & \textbf{A} & \textbf{B} & \textbf{C} \\
  \midrule
  p1 & 1 & 0 & 0 \\
  p2 & 0 & 1 & 0 \\
  p3 & 0 & 0 & 1 \\
  \bottomrule
  \end{tabular}}
  \vspace{-.5em}
\caption{HSE sanity checks. (a) Identical models produce zero
diversity. (b) Orthogonal models achieve maximum diversity
(normalised HSE = 1.0 regardless of society size). p1--p3 represent different prompts.\label{tab:hse-sanity} }
\end{table}

%% file: table/hse_compact2.tex
% ── Table 2: duplicate invariance ───────────────────────────────
\begin{table}[t]
\centering
\label{tab:hse-duplicates}
\small
\begin{tabular}{lcccc}
\toprule
\textbf{Prompt}
  & \textbf{A} & \textbf{B} & \textbf{C}
  & \cellcolor{blue!15}\textbf{D} \\
\midrule
p1 & 1 & 0 & 0 & \cellcolor{blue!15}0 \\
p2 & 0 & 1 & 1 & \cellcolor{blue!15}1 \\
p3 & 1 & 0 & 1 & \cellcolor{blue!15}1 \\
\rowcolor{orange!20}
p4 & 1 & 0 & 1 & \cellcolor{gray!25}{---} \\
\bottomrule
\end{tabular}

\smallskip
\begin{tabular}{lcc}
\toprule
\textbf{Case} & \textbf{HSE} & \textbf{Norm HSE} \\
\midrule
(a) Base: three different models       & 0.654 & 0.413 \\
(b) $+$ duplicate model (D\,=\,C)     & 0.607 & 0.405 \\
(c) $+$ duplicate prompt (p4\,=\,p3)  & 0.610 & 0.385 \\
\bottomrule
\end{tabular}
 \vspace{-.5em}
\caption{HSE on three related cases sharing a common base matrix
  (unshaded $3{\times}3$ block, case~a).
  The \colorbox{blue!15}{shaded column} extends the base with a
  duplicate model D\,=\,C (case~b);
  the \colorbox{orange!20}{shaded row} extends it with a duplicate
  prompt p4\,=\,p3 (case~c).
  Adding duplicates does not increase diversity.
  \label{tab:hse-duplicates}}
\end{table}

%% file: appendix.tex
\appendix
\section{Description of Router Benchmarks}
\label{sec:router:benchmarks}

\subsection{EmbedLLM}
\label{app:embedllm}

EmbedLLM~\citep{zhuang2025embedllm} is a model routing and correctness-forecasting benchmark built around a matrix of binary performance labels collected from 112 open-source language models spanning a wide range of sizes and specialisations (general-purpose, coding, biomedical, physics). Each model was evaluated on 36,054 questions drawn from the test sets of ten established benchmarks, summarised in Table~\ref{tab:embedllm_benchmarks}. For each model--question pair the label records whether the model answered correctly (1) or not (0), yielding a correctness matrix $Y \in \{0,1\}^{112 \times 36{,}054}$. Some questions were duplicates or near-duplicates which differed only by the order in which the multiple choice answers were presented, so we filtered out about 3k prompts. The remaining questions were 
% embedded with the \texttt{all-mpnet-base-v2} sentence transformer (dimension 768) and 
split 80\%/10\%/10\% into train, validation, and test sets (approximately 26K/3.6K/3.6K questions).

In our work, we treat each model as an actor $r_k \in \mathcal{R}$ and each row of $Y$ as the actor's behavioural vector $\mathbf{b}_k$ (Section~\ref{sec:preliminaries}). The full set of 112 models constitutes the \emph{default society}; the \emph{max-HSE subset} is the 9-model subset selected greedily to maximise normalised HSE (Section~\ref{sec:expsetup}).

\begin{table}[t]
\centering
\footnotesize
\caption{Source benchmarks in EmbedLLM with the corresponding domain.}
\label{tab:embedllm_benchmarks}
\begin{tabularx}{\linewidth}{@{}lX@{}}
\toprule
\textbf{Benchmark} & \textbf{Domain} \\
\midrule
MMLU~\citep{hendrycks2021mmlu}         & General knowledge (57 subjects) \\%& 14,042 \\
TruthfulQA~\citep{lin2022truthfulqa}   & Truthfulness / factuality   \\%    & 813   \\
SocialQA~\citep{sap2019socialiqa}      & Social commonsense reasoning   \\% & 1,954  \\
PIQA~\citep{bisk2019piqa}              & Physical commonsense reasoning  \\%& 1,838  \\
MedMCQA~\citep{pal2022medmcqa}         & Medical knowledge              \\ %& 4,183  \\
MathQA~\citep{amini2019mathqa}         & Mathematical word problems     \\ %& 2,985  \\
LogiQA~\citep{liu2020logiqa}           & Logical reasoning              \\% & 651    \\
GSM8K~\citep{cobbe2021gsm8k}           & Grade-school math reasoning    \\% & 1,319  \\
GPQA~\citep{rein2023gpqa}              & Graduate-level science (hard)   \\ %& 448    \\
ASDiV~\citep{miao2020asdiv}            & Arithmetic word problems      \\  %& 2,096  \\
%\midrule
%\textbf{Total}     &                                 & \textbf{30,054} \\
\bottomrule
\end{tabularx}
\end{table}

\subsection{RouterBench}
\label{sec:routerbench}

RouterBench~\citep{hu2024routerbench} is a comprehensive benchmark for evaluating multi-LLM routing systems, designed to cover a broad spectrum of tasks and domains. The dataset contains 405,467 pre-generated inference samples from 11 LLMs --- six open-source (Llama-70B-chat, Mixtral-8x7B, Yi-34B-chat, Code~Llama-34B, Mistral-7B-chat, WizardLM-13B) and five proprietary (GPT-4, GPT-3.5-turbo, Claude-instant-v1, Claude-v1, Claude-v2) --- evaluated across 8 datasets spanning 64~tasks (Table~\ref{tab:routerbench_benchmarks}).

Each sample records the model response alongside a quality score and an inference cost in dollars, enabling joint evaluation of task performance and economic cost. For routing experiments, the dataset is partitioned  80\%/10\%/10\% into train, validation, and test sets (approximately 29K/3.6K/3.6K prompts). 
% A separate RAG dataset of 800 real user queries, evaluated on 14 models (the 11 above plus three online-search models: You.com, sonar-small-online, sonar-medium-online), is included to assess routing in retrieval-augmented settings.

In our work, we treat each of the 11 models as an actor $r_k \in \mathcal{R}$, using binary correctness labels as the entries of the behavioural matrix $B$ (Section~\ref{sec:preliminaries}). The full set of 11 models constitutes the \emph{default society}; the \emph{max-HSE subset} is the 4-model subset that maximises normalised HSE, selected greedily (Section~\ref{sec:expsetup}).

\begin{table}[t]
\centering
\caption{Datasets included in RouterBench, grouped by task category. MMLU accounts for 57 of the 64 tasks; the remaining seven datasets each contribute one task.}
\label{tab:routerbench_benchmarks}
\small
\begin{tabularx}{\linewidth}{@{}lX@{}}
\toprule
\textbf{Benchmark} & \textbf{Domain} \\
\midrule
HellaSwag~\citep{zellers2019hellaswag}  & Commonsense Reasoning    \\
Winogrande~\citep{sakaguchi2021winogrande} & Commonsense Reasoning \\
ARC Challenge~\citep{clark2018arc}  & Commonsense Reasoning        \\
 MMLU~\citep{hendrycks2021mmlu}           & Knowledge Understanding  \\
MT-Bench~\citep{zheng2023mtbench}          & Conversation         \\
 GSM8K~\citep{cobbe2021gsm8k}               & Math                   \\
MBPP~\citep{austin2021mbpp}                & 1 Coding               \\
% Martian RAG (proprietary)                  & 1 RAG                \\

\bottomrule
\end{tabularx}
\end{table}

\subsection{Synthetic Societies}
\label{sec:synthetic}

\paragraph{Subject-Area Clustering (SA)}
To construct the SA synthetic society we assign each of the 67 evaluation items in EmbedLLM,  comprising 57 MMLU subjects and 10 additional benchmarks (GSM8K, MathQA, ASDiv, LogiQA, GPQA, MedMCQA, TruthfulQA, SocialQA, PIQA),  to one of 15 mutually exclusive subject-area clusters. The assignment criterion is academic domain: items are grouped by what they are \emph{about}, independently of how much reasoning they require. The resulting clusters are shown in Table~\ref{tab:sa-clusters}.

\begin{table*}[t]
\centering
\small
\setlength{\tabcolsep}{5pt}
\renewcommand{\arraystretch}{1.12}
\begin{tabular}{@{}l p{5.2cm} p{6.2cm} p{1.5cm} @{}}
\toprule
\textbf{ID} & \textbf{Cluster} & \textbf{MMLU subjects} & \textbf{EmbedLLM} \\
\midrule
C01 & Elementary \& school arithmetic
    & Elementary mathematics, HS statistics
    & GSM8K, ASDiv \\
C02 & Advanced \& university mathematics
    & Abstract algebra, college mathematics, HS mathematics, econometrics
    & MathQA \\
C03 & Logic, argumentation \& formal reasoning
    & Formal logic, logical fallacies
    & LogiQA  \\
C04 & Physics \& astronomy
    & Astronomy, college physics, conceptual physics, HS physics
    & ---  \\
C05 & Chemistry \& earth sciences
    & College chemistry, HS chemistry
    & GPQA  \\
C06 & Biology, genetics \& virology
    & College biology, HS biology, medical genetics, virology
    & ---  \\
C07 & Clinical medicine \& healthcare
    & Anatomy, clinical knowledge, college medicine, human aging, human
      sexuality, nutrition, pre-medical, professional medicine
    & MedMCQA \\
C08 & Computer science, AI \& cybersecurity
    & College CS, computer security, electrical engineering, HS CS,
      machine learning
    & ---\\
C09 & Economics, business \& accounting
    & HS macroeconomics, HS microeconomics, management, marketing,
      professional accounting, public relations
    & ---  \\
C10 & Law \& legal reasoning
    & Business ethics, international law, jurisprudence, pre-law,
      professional law
    & ---  \\
C11 & Ethics, philosophy \& religion
    & Moral disputes, moral scenarios, philosophy, world religions
    & ---\\
C12 & Psychology \& behavioural sciences
    & HS psychology, professional psychology, sociology
    & SocialQA, TruthfulQA  \\
C13 & Politics, international relations \& security
    & HS government \& politics, security studies, US foreign policy
    & ---  \\
C14 & History, geography \& world knowledge
    & Global facts, HS European history, HS geography, HS US history,
      HS world history, miscellaneous
    & ---  \\
C15 & Commonsense \& physical intuition
    & ---
    & PIQA \\
\bottomrule
\end{tabular}
\caption{Subject-area clustering of the 67 EmbedLLM evaluation items  into 15
  groups. The \textbf{MMLU subjects} column lists the 57 individual
  MMLU subjects assigned to each cluster; the \textbf{EmbedLLM}
  column lists the 10 additional benchmarks. Each synthetic SA expert
  scores 1 on every item in its cluster and 0 on all others.
  HS~=~High School; CS~=~Computer Science.}
\label{tab:sa-clusters}
\end{table*}

Each synthetic SA agent is assigned a binary performance vector, scoring 1 on every item in its cluster and 0 on all others, modelling a specialist with perfect but narrowly scoped expertise. Several borderline assignments are worth noting. Econometrics is placed in Advanced Mathematics rather than Economics because its questions are dominated by statistical modelling and regression rather than economic theory. GPQA is placed in Chemistry \& Earth Sciences because its hardest and most distinctive questions are in graduate-level chemistry and physical sciences. TruthfulQA is placed in Psychology \& Behavioural Sciences because its core demandm  resisting socially transmitted false beliefs, is an epistemic and psychological phenomenon rather than a factual one. Business Ethics is placed in Law \& Legal Reasoning because its MMLU questions predominantly test corporate governance and fiduciary duty rather than philosophical ethics.

\paragraph{Reasoning-Depth Clustering (RD)}
To construct the RD synthetic society we assign the same 67 EmbedLLM items to one of five reasoning-depth tiers that form a strictly ordered scale from surface recall to adversarial expert reasoning. The assignment criterion is the dominant cognitive operation required to answer correctly, independent of subject matter. 

\begin{description}
\item[\textbf{R1 (Direct Lookup)}] covers items answerable by retrieving a single stored fact with no inference, such as recalling a historical date or a viral property (13 items). 

\item[\textbf{R2 (Concept Application)}] covers items requiring one inferential step — applying a rule, definition, or principle to a specific case — including PIQA, SocialQA, and TruthfulQA (21 items). 

\item[\textbf{R3 (Multi-Step Reasoning)}] covers items whose solution requires composing two or more dependent inference steps, including GSM8K, MathQA, and MedMCQA (21 items).

\item[\textbf{R4 (Formal Symbolic Reasoning)}] covers items solved by mechanically applying a formal system such as logic, algebra, or proof calculus, including LogiQA and ASDiv (7 items). 

\item[\textbf{R5 (Adversarial \& Expert-Level)}] covers items specifically designed to defeat shallow heuristics, requiring deep domain expertise and resistance to carefully crafted near-miss distractors, including GPQA and the four MMLU Professional subjects (5 items). 
\end{description}

As with SA, each RD agent scores 1 on every item in its tier and 0 elsewhere. Key borderline decisions include: TruthfulQA is assigned to R2 rather than R3 because each individual question requires a single evaluative judgment, not a reasoning chain; LogiQA is assigned to R4 rather than R3 because its civil-service puzzles require applying formal inference rules rather than multi-step domain reasoning; ASDiv is assigned to R4 rather than R1 because its design goal is robustness across diverse arithmetic structures, probing rule application rather than recall; and GPQA is assigned to R5 because it is explicitly designed so that non-experts perform near chance and domain experts reach only around 65\%, with distractors crafted to exploit common expert heuristics.

\section{Query perturbations}

\subsection{Perturbation generation}
\label{app:perturbation_templates}

We generate the query perturbations by prompting an LM, using Gemini 3.1 Flash-Lite \cite{pichai2025new} as backbone model. This section documents the five zero-shot prompts used to generate
perturbations of each query in the evaluation sets of EmbedLLM and RouterBench (Section~\ref{sec:perturbations}).
Each prompt instructs the model to produce 10 rewrites of an input query
\pvar{prompt} according to specific perturbation instructions for each of the five
levels: character, word, syntax, paraphrase, and
rambling. We request more perturbations than needed since occasionally the model produces duplicates variants. We then parse the outputs and select 5 unique variants per perturbation category. 

%% ════════════════════════════════════════════════════════════════════════════

\begin{tcolorbox}[prompt box,
  title={\small\bfseries\color{white} Level 1: Character-Level Perturbations},
  attach boxed title to top left={yshift=-2mm, xshift=6mm},
  boxed title style={header box},
  title after break={\small\bfseries\color{white} Level 1 (cont.)},
]

Please write 10 different rewrites (perturbations) of the prompt provided
below by introducing character-level perturbations such as inserting typos,
mimicking a user typing quickly on a keyboard, for instance hitting adjacent
keys, or replacing a character with one adjacent to it (e.g., `gello'
instead of `hello'), adding or removing random characters (e.g., `thng'
instead of `thing'), swapping two adjacent letters (e.g., `ehllo' instead of
`hello'), or using characters that look similar (e.g.,
`b3st' instead of `best', or `fr\'iend' instead of `friend').

\medskip
\begin{tcolorbox}[field header] Output Format \end{tcolorbox}

Format your answer by numbering each rewrite as 1., 2., 3.\ etc.:

\begin{tcolorbox}[output box]
1. first rewrite\\
2. second rewrite\\
\ldots
\end{tcolorbox}

\medskip
\begin{tcolorbox}[field header] Prompt \end{tcolorbox}
\pvar{prompt}

\end{tcolorbox}

\medskip
%% ════════════════════════════════════════════════════════════════════════════

\begin{tcolorbox}[prompt box,
  title={\small\bfseries\color{white} Level 2: Word-Level Perturbations},
  attach boxed title to top left={yshift=-2mm, xshift=6mm},
  boxed title style={header box},
  title after break={\small\bfseries\color{white} Level 2 (cont.)},
]

Please write 10 different rewrites (perturbations) of the prompt provided
below by introducing word-level perturbations, while trying to keep the
original meaning (semantics) intact, such as replacing a keyword with a
synonym (e.g., `closest friend' instead of `best friend'), misspelling
certain words or writing them phonetically, or deleting a non-keyword such
as ``the,'' ``is,'' or ``of.''

\medskip
\begin{tcolorbox}[field header] Output Format \end{tcolorbox}

Format your answer by numbering each rewrite as 1., 2., 3.\ etc.:

\begin{tcolorbox}[output box]
1. first rewrite\\
2. second rewrite\\
\ldots
\end{tcolorbox}

\medskip
\begin{tcolorbox}[field header] Prompt \end{tcolorbox}
\pvar{prompt}

\end{tcolorbox}

\medskip
%% ════════════════════════════════════════════════════════════════════════════

\begin{tcolorbox}[prompt box,
  title={\small\bfseries\color{white} Level 3: Syntax-Level Perturbations},
  attach boxed title to top left={yshift=-2mm, xshift=6mm},
  boxed title style={header box},
  title after break={\small\bfseries\color{white} Level 3 (cont.)},
]

Please write 10 different rewrites (perturbations) of the prompt provided
below by introducing sentence-level perturbations, which alter the structure
or grammar of the prompt without changing the core meaning.

\medskip
\begin{tcolorbox}[field header] Output Format \end{tcolorbox}

Format your answer by numbering each rewrite as 1., 2., 3.\ etc.:

\begin{tcolorbox}[output box]
1. first rewrite\\
2. second rewrite\\
\ldots
\end{tcolorbox}

\medskip
\begin{tcolorbox}[field header] Prompt \end{tcolorbox}
\pvar{prompt}

\end{tcolorbox}

\medskip
%% ════════════════════════════════════════════════════════════════════════════

\begin{tcolorbox}[prompt box,
  title={\small\bfseries\color{white} Level 4: Paraphrase Perturbations},
  attach boxed title to top left={yshift=-2mm, xshift=6mm},
  boxed title style={header box},
  title after break={\small\bfseries\color{white} Level 4 (cont.)},
]

Please write 10 different rewrites (perturbations) of the prompt provided
below by paraphrasing without changing the core meaning, for instance by
re-writing the prompt in a different voice, style, or tone.

\medskip
\begin{tcolorbox}[field header] Output Format \end{tcolorbox}

Format your answer by numbering each rewrite as 1., 2., 3.\ etc.:

\begin{tcolorbox}[output box]
1. first rewrite\\
2. second rewrite\\
\ldots
\end{tcolorbox}

\medskip
\begin{tcolorbox}[field header] Prompt \end{tcolorbox}
\pvar{prompt}

\end{tcolorbox}

\medskip
%% ════════════════════════════════════════════════════════════════════════════

\begin{tcolorbox}[prompt box,
  title={\small\bfseries\color{white} Level 5: Rambling Perturbations},
  attach boxed title to top left={yshift=-2mm, xshift=6mm},
  boxed title style={header box},
  title after break={\small\bfseries\color{white} Level 5 (cont.)},
]

Please write 10 different rewrites (perturbations) of the prompt provided
below by re-writing the entire prompt but with some unrelated noise before
or after the prompt, without changing the core meaning.

\medskip
\begin{tcolorbox}[field header] Output Format \end{tcolorbox}

Format your answer by numbering each rewrite as 1., 2., 3.\ etc.:

\begin{tcolorbox}[output box]
1. first rewrite\\
2. second rewrite\\
\ldots
\end{tcolorbox}

\medskip
\begin{tcolorbox}[field header] Prompt \end{tcolorbox}
\pvar{prompt}

\end{tcolorbox}

\subsection{Query perturbation examples}

% ─────────────────────────────────────────────────────────────────────────────
% Required in the preamble (if not already defined):
%   \usepackage{tcolorbox}
%   \usepackage{enumitem}
%   \tcbuselibrary{skins,breakable}
%
%   \definecolor{promptbg}{HTML}{F7F7F8}
%   \definecolor{promptframe}{HTML}{4A90D9}
%   \definecolor{headerblue}{HTML}{4A90D9}
%   \definecolor{fieldbg}{HTML}{E8E8E8}
%   \definecolor{outputbg}{HTML}{F0F0F0}
%
\tcbset{
example box/.style={
  enhanced, breakable, colback=promptbg, colframe=promptframe,
  fontupper=\small, left=4mm, right=4mm, top=4mm, bottom=3mm,
  arc=1mm, boxrule=0.5pt
},
dataset header/.style={
  colback=headerblue, colframe=headerblue, arc=1mm, boxrule=0pt
},
field header/.style={
  colback=fieldbg, colframe=fieldbg, arc=0.5mm, boxrule=0pt,
  fontupper=\small\bfseries, left=2mm, right=2mm,
  top=0.5mm, bottom=0.5mm
},
content box/.style={
  colback=outputbg, colframe=outputbg, arc=0.5mm, boxrule=0pt,
  fontupper=\small\ttfamily, left=2mm, right=2mm,
  top=1mm, bottom=1mm
},
}
% ─────────────────────────────────────────────────────────────────────────────

This section presents some examples of the generated prompt perturbations.

%% ════════════════════════════════════════════════════════════════════════════
\medskip

% --- EMBEDLLM PROMPT 1 ---
\begin{tcolorbox}[example box,
  title={\small\bfseries\color{white} EmbedLLM Perturbation Example 1},
  attach boxed title to top left={yshift=-2mm, xshift=6mm},
  boxed title style={dataset header},
]
\begin{tcolorbox}[field header] Original Prompt \end{tcolorbox}
\begin{tcolorbox}[content box]
How to reduce the pain of a sore throat with a natural substance?
\end{tcolorbox}

\medskip
\begin{tcolorbox}[field header] Perturbation Variants \end{tcolorbox}
\begin{description}[leftmargin=1em, style=nextline, itemsep=4pt, parsep=0pt, topsep=4pt]
  \item[\textbf{Level 1 (Character):}]
    \texttt{How to reduce the pain of a sore throat wth a natural subtsance}
  \item[\textbf{Level 2 (Word):}]
    \texttt{Alleviate pain from a sore throat using a natural remedy}
  \item[\textbf{Level 3 (Syntax):}]
    \texttt{How can I use a natural substance to reduce the pain of a sore? }
  \item[\textbf{Level 4 (Paraphrase):}]
    \texttt{What are some natural remedies I can use to soothe a sore throat?}
  \item[\textbf{Level 5 (Rambling):}]
    \texttt{The sky is blue today. Question: How to reduce the pain of a sore throat with a natural substance?}
\end{description}
\end{tcolorbox}

\medskip

% --- EMBEDLLM PROMPT 2 ---
\begin{tcolorbox}[example box,
  title={\small\bfseries\color{white} EmbedLLM Perturbation Example 2},
  attach boxed title to top left={yshift=-2mm, xshift=6mm},
  boxed title style={dataset header},
]
\begin{tcolorbox}[field header] Original Prompt \end{tcolorbox}
\begin{tcolorbox}[content box]
How can I get something sweet even when I am not eating sugar?
\end{tcolorbox}

\medskip
\begin{tcolorbox}[field header] Perturbation Variants \end{tcolorbox}
\begin{description}[leftmargin=1em, style=nextline, itemsep=4pt, parsep=0pt, topsep=4pt]
  \item[\textbf{Level 1 (Character):}]
    \texttt{Hoq can I get somethin gswwet even when I am not eating suagr?}
  \item[\textbf{Level 2 (Word):}]
    \texttt{How can I satisfy a sweet tooth if I am not consuming sugar?}
  \item[\textbf{Level 3 (Syntax):}]
    \texttt{If I am abstaining from sugar, how can I still enjoy something sweet?}
  \item[\textbf{Level 4 (Paraphrase):}]
    \texttt{How can I satisfy a craving for sweetness without consuming any sugar?}
  \item[\textbf{Level 5 (Rambling):}]
    \texttt{System status: online. How can I get something sweet even when I am not eating sugar?}
\end{description}
\end{tcolorbox}

%% ════════════════════════════════════════════════════════════════════════════

\medskip

% --- ROUTERBENCH PROMPT 1 ---
\begin{tcolorbox}[example box,
  title={\small\bfseries\color{white} RouterBench Perturbation Example 1},
  attach boxed title to top left={yshift=-2mm, xshift=6mm},
  boxed title style={dataset header},
]
\begin{tcolorbox}[field header] Original Prompt \end{tcolorbox}
\begin{tcolorbox}[content box]
Write a function to find the n'th lucas number.
\end{tcolorbox}

\medskip
\begin{tcolorbox}[field header] Perturbation Variants \end{tcolorbox}
\begin{description}[leftmargin=1em, style=nextline, itemsep=4pt, parsep=0pt, topsep=4pt]
  \item[\textbf{Level 1 (Character):}]
    \texttt{Wrtie a function to find the n'th lucas number.}
  \item[\textbf{Level 2 (Word):}]
    \texttt{Create a function to find the n'th lucas number.}
  \item[\textbf{Category C (Syntax):}]
    \texttt{Create a function that calculates the nth Lucas number.}
  \item[\textbf{Level 4 (Paraphrase):}]
    \texttt{Please code a solution to determine the Lucas number at index n.}
  \item[\textbf{Level 5 (Rambling):}]
    \texttt{It’s been a long day. Write a function to find the n'th lucas number.}
\end{description}
\end{tcolorbox}

\medskip

% --- ROUTERBENCH PROMPT 2 ---
\begin{tcolorbox}[example box,
  title={\small\bfseries\color{white} RouterBench Perturbation Example 2},
  attach boxed title to top left={yshift=-2mm, xshift=6mm},
  boxed title style={dataset header},
]
\begin{tcolorbox}[field header] Original Prompt \end{tcolorbox}
\begin{tcolorbox}[content box]
Write a function to check if a dictionary is empty
\end{tcolorbox}

\medskip
\begin{tcolorbox}[field header] Perturbation Variants \end{tcolorbox}
\begin{description}[leftmargin=1em, style=nextline, itemsep=4pt, parsep=0pt, topsep=4pt]
  \item[\textbf{Level 1 (Character):}]
    \texttt{Writ a function to check if a dictionary is empty}
  \item[\textbf{Level 2 (Word):} ]
    \texttt{Create a function to verify if a dict is empty}
  \item[\textbf{Level 3 (Syntax):} ]
    \texttt{Create a function that verifies whether a dictionary is empty.}
  \item[\textbf{Level 4 (Paraphrase):} ]
    \texttt{Create a function that determines if a dictionary contains no elements.}
  \item[\textbf{Level 5 (Rambling):}]
    \texttt{The weather is nice today. Please write a function to check if a dictionary is empty.}
\end{description}
\end{tcolorbox}
% ============================================================

\section{Prompted Router Policy}
\label{app:prompted_router}

For the prompted router policy, we use a few-shot prompt template with Gemini 3.1 Flash-Lite \cite{pichai2025new} as LM backbone. 
The prompt asks a language model to select the best-suited actor for a given query, conditioned on (i)~a natural-language
description of each available actor, and (ii)~a set of
few-shot exemplars drawn from the training split. 

Section~\ref{app:prompted_router_template}
presents the template itself; Section~\ref{app:exemplar_selection}
describes how few-shot exemplars are selected;
and Section~\ref{app:prompted_instantiation}
show how the template is instantiated for the EmbedLLM and RouterBench datasets.

%% ════════════════════════════════════════════════════════════════════════════
\subsection{Prompt Template}
\label{app:prompted_router_template}

\begin{tcolorbox}[prompt box,
  title={\small\bfseries\color{white} Few-Shot Routing Prompt Template},
  attach boxed title to top left={yshift=-2mm, xshift=6mm},
  boxed title style={header box},
  title after break={\small\bfseries\color{white} Prompt Template (cont.)},
]

\begin{tcolorbox}[field header] Instructions \end{tcolorbox}

Given a prompt, select which model is best positioned to
answer.

Your answer should be one of: \pvar{actors}.

\medskip
\begin{tcolorbox}[field header] Context \end{tcolorbox}

The following models are available:

\pvar{actors\_description}

\medskip
\begin{tcolorbox}[field header] Few-Shot Exemplars \end{tcolorbox}

\pvar{few\_shot\_examples}

\medskip
\begin{tcolorbox}[field header] Prompt \end{tcolorbox}

PROMPT: \pvar{prompt}

Best model for this prompt:
\end{tcolorbox}

%% ════════════════════════════════════════════════════════════════════════════
\subsection{Few-Shot Exemplar Selection}
\label{app:exemplar_selection}

For each actor, $k$ exemplars are randomly sampled from training
prompts where that actor is the clear winner, i.e., its score
exceeds all other actors by at least a margin
$\delta$.  We use $k{=}4$ and $\delta{=}0.4$. In the rare instances in which there are less than $k$ prompts that satisfy this criteria, we fall back to sampling from training prompts where that actor achieved the highest score among all actors in the society (possibly tied with other actors). 

Exemplars are formatted as below and shuffled together across all
specialists:

\begin{tcolorbox}[output box]
PROMPT: \pvar{exemplar\_query}\\
Best model for this prompt: \pvar{actor}
\end{tcolorbox}

%% ════════════════════════════════════════════════════════════════════════════
\subsection{Instantiation}
\label{app:prompted_instantiation}

For EmbedLLM and RouterBench we list all the names of the LLMs available in the dataset, e.g., \texttt{huggingfaceh4\_zephyr\_7b\_beta},  in the \pvar{actors\_description} field.

For the \emph{synthetic} societies, each actor
corresponds to a knowledge domain (subject area), a reasoning depth,
or both, so we inject the descriptions below into the
\pvar{actors\_description} placeholder.

\medskip

\begin{tcolorbox}[prompt box,
  title={\small\bfseries\color{white} EmbedLLM Synthetic SA},
  attach boxed title to top left={yshift=-2mm, xshift=6mm},
  boxed title style={header box},
  title after break={\small\bfseries\color{white} Subject Areas (cont.)},
]

\begin{itemize}[leftmargin=1.5em, itemsep=1pt, parsep=0pt]
  \item \texttt{elementary\_math}: a specialist in foundational numerical
    operations and word problems at primary/middle-school level
  \item \texttt{advanced\_math}: a specialist in university-level pure and
    applied mathematics requiring symbolic manipulation, proof, or
    multi-domain quantitative reasoning
  \item \texttt{logic}: a specialist in the application of formal logical
    rules, identification of valid/invalid argument structures, and
    detection of informal fallacies
  \item \texttt{physics}: a specialist in principles of mechanics,
    electromagnetism, thermodynamics, quantum physics, and astrophysics at
    high-school to graduate level
  \item \texttt{chemistry}: a specialist in chemical principles from
    stoichiometry and bonding to organic reactions and physical chemistry,
    plus graduate-level natural science breadth
  \item \texttt{biology}: a specialist in life sciences spanning molecular
    biology, genetics, evolutionary biology, ecology, and virology at school
    to university level
  \item \texttt{medicine}: a specialist in clinical diagnosis, patient
    management, pharmacology, and healthcare practice at the level of
    medical school and professional licensing
  \item \texttt{computer\_science}: a specialist in algorithms, data
    structures, machine learning models, computer systems, and
    security---theoretical and applied CS at school to graduate level
  \item \texttt{economics}: a specialist in micro/macroeconomic theory,
    business management, marketing, public relations, and professional
    accounting
  \item \texttt{law}: a specialist in legal theory, jurisprudence,
    international and domestic law at the level of law school entry and
    professional bar examination
  \item \texttt{philosophy}: a specialist in normative ethics, applied moral
    reasoning, philosophy of mind and knowledge, and comparative religion
  \item \texttt{psychology}: a specialist in cognitive, developmental,
    clinical, and social psychology, plus professional assessment and
    therapy practice
  \item \texttt{politics}: a specialist in political institutions,
    government, foreign policy, international security, and strategic
    studies
  \item \texttt{history}: a specialist in historical events and movements
    (European, US, world), human and physical geography, and broad
    cross-national factual knowledge
  \item \texttt{commonsense}: a specialist in implicit everyday knowledge
    about how the physical world works and how people behave
\end{itemize}

\end{tcolorbox}

\medskip

\begin{tcolorbox}[prompt box,
  title={\small\bfseries\color{white} EmbedLLM Synthetic RD},
  attach boxed title to top left={yshift=-2mm, xshift=6mm},
  boxed title style={header box},
  title after break={\small\bfseries\color{white} Reasoning Depths (cont.)},
]

\begin{itemize}[leftmargin=1.5em, itemsep=1pt, parsep=0pt]
  \item \texttt{direct\_lookup}: a model whose answers are stored facts
    retrievable in one step, with no inference, calculation, or reasoning
    chain needed
  \item \texttt{concept\_application}: a model that applies a concept,
    rule, or definition to a specific case in one inferential step,
    requiring understanding beyond rote recall but no chaining
  \item \texttt{multi\_step}: a model that reaches the answer by composing
    two or more distinct reasoning steps, each depending on the output of
    the previous
  \item \texttt{formal\_symbolic}: a model that derives the answer by
    applying a formal system (logic, algebra, proof) according to explicit
    syntactic rules
  \item \texttt{adversarial}: a model tested on questions specifically
    designed to defeat shallow heuristics, requiring deep domain expertise,
    resistance to plausible distractors, and often meta-cognitive awareness
\end{itemize}

\end{tcolorbox}

\section{KNN Router Policy}
\label{app:knn_router}
The KNN-based router computes an embedding $e(q)$ of the query using Gemini Embedding 2 as model backbone~\citep{lee2025gemini} and retrieves the $k$ nearest neighbours $\mathcal{N}_k(x)$ from the training set using cosine similarity in the embedding space. It then assigns the actor with highest score over the $k$~nearest neighbours. 

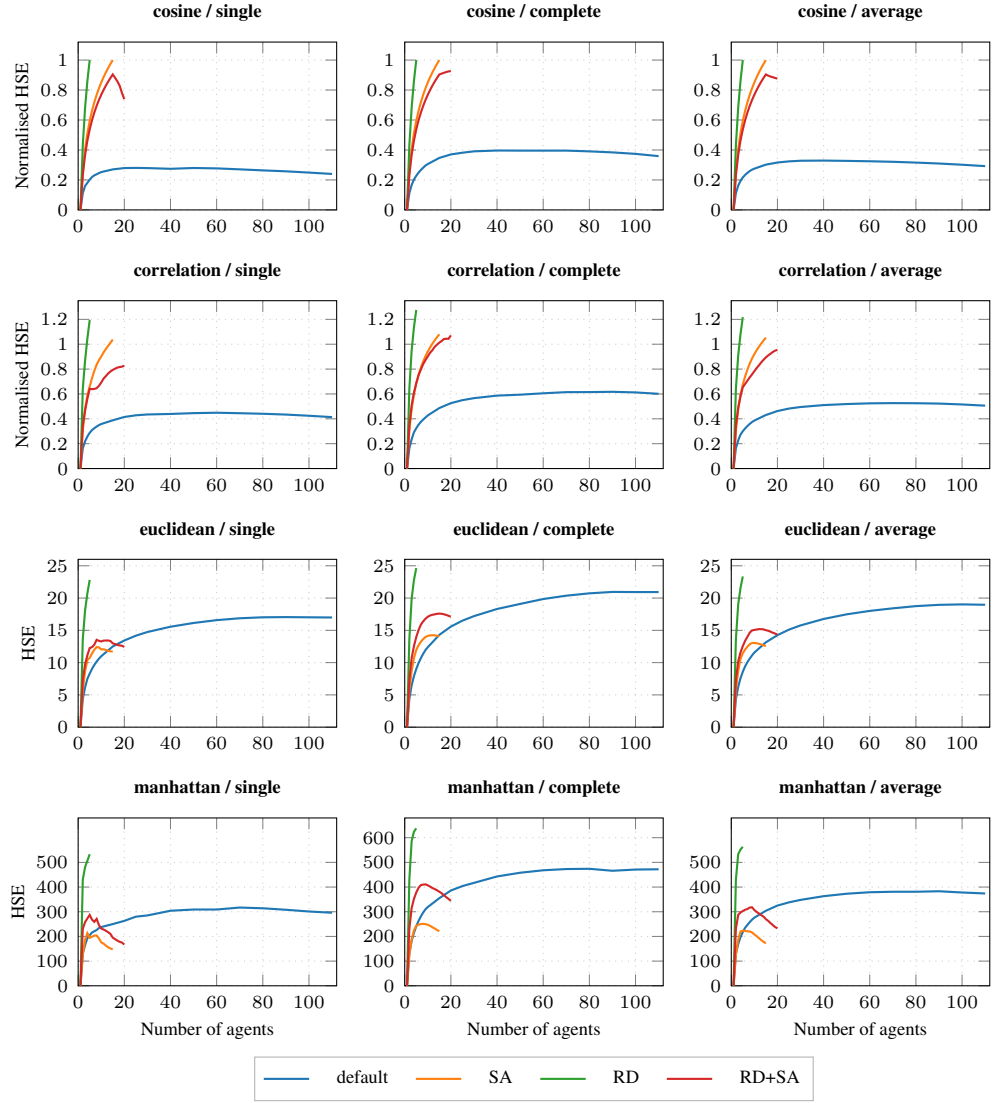
\begin{figure*}[t]
\centering
\pgfplotsset{
  sweep base/.style={
    width=5.0cm, height=3.8cm,
    xmin=0, xmax=112,
    xtick={0,20,40,60,80,100},
    xticklabel style={font=\scriptsize},
    yticklabel style={font=\scriptsize,
      /pgf/number format/fixed,
      /pgf/number format/precision=1},
    xlabel={Number of agents},
    xlabel style={font=\scriptsize},
    xlabel shift=-2pt,
    ylabel style={font=\scriptsize},
    ylabel shift=-3pt,
    title style={font=\scriptsize\bfseries, yshift=-2pt},
    grid=major,
    grid style={dotted, gray!40},
    every axis plot/.append style={thick, mark size=0pt},
  },
}
\begin{tikzpicture}
\begin{groupplot}[
  group style={
    group size=3 by 4,
    horizontal sep=0.9cm,
    vertical sep=1.2cm,
    ylabels at=edge left,
    xlabels at=edge bottom,
  },
  sweep base,
]

%% ════════════════════════════════════════════════════════════
%% ROW 1: cosine  (ymax = 1.1)
%% ════════════════════════════════════════════════════════════

\nextgroupplot[title={cosine / single},
  ylabel={Normalised HSE},
  ymin=0, ymax=1.12,
  ytick={0,0.2,0.4,0.6,0.8,1.0},
  legend to name=sweeplegend,
  legend columns=4,
  legend style={font=\scriptsize, draw=gray!60, fill=white,
    inner sep=3pt, column sep=8pt},
  legend cell align=left,
]
\addplot[mpblue]   coordinates {(1,0.0000)(2,0.1117)(3,0.1602)(4,0.1808)(5,0.2021)(6,0.2194)(7,0.2307)(8,0.2380)(9,0.2454)(10,0.2524)(15,0.2705)(20,0.2792)(25,0.2803)(30,0.2791)(40,0.2743)(50,0.2797)(60,0.2771)(70,0.2707)(80,0.2638)(90,0.2576)(100,0.2491)(110,0.2400)};
\addlegendentry{default}
\addplot[mporange] coordinates {(1,0.0000)(2,0.2560)(3,0.4057)(4,0.5119)(5,0.5943)(6,0.6616)(7,0.7186)(8,0.7679)(9,0.8114)(10,0.8503)(11,0.8855)(12,0.9176)(13,0.9472)(14,0.9745)(15,1.0000)};
\addlegendentry{SA}
\addplot[mpgreen]  coordinates {(1,0.0000)(2,0.4307)(3,0.6826)(4,0.8614)(5,1.0000)};
\addlegendentry{RD}
\addplot[mpred]    coordinates {(1,0.0000)(2,0.2314)(3,0.3667)(4,0.4628)(5,0.5372)(6,0.5981)(7,0.6496)(8,0.6941)(9,0.7335)(10,0.7686)(11,0.8004)(12,0.8295)(13,0.8562)(14,0.8809)(15,0.9040)(16,0.8817)(17,0.8578)(18,0.8274)(19,0.7793)(20,0.7389)};
\addlegendentry{RD+SA}

\nextgroupplot[title={cosine / complete},
  ymin=0, ymax=1.12, ytick={0,0.2,0.4,0.6,0.8,1.0}]
\addplot[mpblue]   coordinates {(1,0.0000)(2,0.1117)(3,0.1661)(4,0.2031)(5,0.2277)(6,0.2492)(7,0.2665)(8,0.2816)(9,0.2964)(10,0.3061)(15,0.3470)(20,0.3703)(25,0.3817)(30,0.3909)(40,0.3964)(50,0.3957)(60,0.3956)(70,0.3957)(80,0.3908)(90,0.3839)(100,0.3740)(110,0.3589)};
\addplot[mporange] coordinates {(1,0.0000)(2,0.2560)(3,0.4057)(4,0.5119)(5,0.5943)(6,0.6616)(7,0.7186)(8,0.7679)(9,0.8114)(10,0.8503)(11,0.8855)(12,0.9176)(13,0.9472)(14,0.9745)(15,1.0000)};
\addplot[mpgreen]  coordinates {(1,0.0000)(2,0.4307)(3,0.6826)(4,0.8614)(5,1.0000)};
\addplot[mpred]    coordinates {(1,0.0000)(2,0.2314)(3,0.3667)(4,0.4628)(5,0.5372)(6,0.5981)(7,0.6496)(8,0.6941)(9,0.7335)(10,0.7686)(11,0.8004)(12,0.8295)(13,0.8562)(14,0.8809)(15,0.9040)(16,0.9094)(17,0.9152)(18,0.9200)(19,0.9245)(20,0.9263)};

\nextgroupplot[title={cosine / average},
  ymin=0, ymax=1.12, ytick={0,0.2,0.4,0.6,0.8,1.0}]
\addplot[mpblue]   coordinates {(1,0.0000)(2,0.1117)(3,0.1615)(4,0.1937)(5,0.2162)(6,0.2337)(7,0.2470)(8,0.2595)(9,0.2685)(10,0.2753)(15,0.3022)(20,0.3163)(25,0.3239)(30,0.3281)(40,0.3292)(50,0.3272)(60,0.3246)(70,0.3211)(80,0.3158)(90,0.3096)(100,0.3015)(110,0.2922)};
\addplot[mporange] coordinates {(1,0.0000)(2,0.2560)(3,0.4057)(4,0.5119)(5,0.5943)(6,0.6616)(7,0.7186)(8,0.7679)(9,0.8114)(10,0.8503)(11,0.8855)(12,0.9176)(13,0.9472)(14,0.9745)(15,1.0000)};
\addplot[mpgreen]  coordinates {(1,0.0000)(2,0.4307)(3,0.6826)(4,0.8614)(5,1.0000)};
\addplot[mpred]    coordinates {(1,0.0000)(2,0.2314)(3,0.3667)(4,0.4628)(5,0.5372)(6,0.5981)(7,0.6496)(8,0.6941)(9,0.7335)(10,0.7686)(11,0.8004)(12,0.8295)(13,0.8562)(14,0.8809)(15,0.9040)(16,0.8959)(17,0.8900)(18,0.8856)(19,0.8802)(20,0.8764)};

%% ════════════════════════════════════════════════════════════
%% ROW 2: correlation  (ymax ~1.35)
%% ════════════════════════════════════════════════════════════

\nextgroupplot[title={correlation / single},
  ylabel={Normalised HSE},
  ymin=0, ymax=1.35, ytick={0,0.2,0.4,0.6,0.8,1.0,1.2}]
\addplot[mpblue]   coordinates {(1,0.0000)(2,0.1628)(3,0.2189)(4,0.2583)(5,0.2876)(6,0.3093)(7,0.3246)(8,0.3376)(9,0.3492)(10,0.3582)(15,0.3881)(20,0.4148)(25,0.4282)(30,0.4350)(40,0.4391)(50,0.4457)(60,0.4489)(70,0.4451)(80,0.4406)(90,0.4341)(100,0.4246)(110,0.4135)};
\addplot[mporange] coordinates {(1,0.0000)(2,0.3048)(3,0.4768)(4,0.5917)(5,0.6601)(6,0.7280)(7,0.7860)(8,0.8353)(9,0.8721)(10,0.9009)(11,0.9337)(12,0.9615)(13,0.9872)(14,1.0126)(15,1.0371)};
\addplot[mpgreen]  coordinates {(1,0.0000)(2,0.6401)(3,0.8693)(4,1.0466)(5,1.1947)};
\addplot[mpred]    coordinates {(1,0.0000)(2,0.3439)(3,0.4670)(4,0.5623)(5,0.6418)(6,0.6385)(7,0.6405)(8,0.6462)(9,0.6656)(10,0.6919)(11,0.7234)(12,0.7432)(13,0.7649)(14,0.7808)(15,0.7934)(16,0.8032)(17,0.8122)(18,0.8162)(19,0.8197)(20,0.8272)};

\nextgroupplot[title={correlation / complete},
  ymin=0, ymax=1.35, ytick={0,0.2,0.4,0.6,0.8,1.0,1.2}]
\addplot[mpblue]   coordinates {(1,0.0000)(2,0.1628)(3,0.2352)(4,0.2917)(5,0.3241)(6,0.3529)(7,0.3756)(8,0.3940)(9,0.4106)(10,0.4274)(15,0.4867)(20,0.5260)(25,0.5501)(30,0.5661)(40,0.5867)(50,0.5941)(60,0.6052)(70,0.6145)(80,0.6149)(90,0.6173)(100,0.6122)(110,0.6012)};
\addplot[mporange] coordinates {(1,0.0000)(2,0.3048)(3,0.4774)(4,0.5976)(5,0.6858)(6,0.7546)(7,0.8117)(8,0.8608)(9,0.9035)(10,0.9409)(11,0.9742)(12,1.0042)(13,1.0315)(14,1.0567)(15,1.0801)};
\addplot[mpgreen]  coordinates {(1,0.0000)(2,0.6401)(3,0.9488)(4,1.1364)(5,1.2753)};
\addplot[mpred]    coordinates {(1,0.0000)(2,0.3439)(3,0.5097)(4,0.6105)(5,0.6852)(6,0.7511)(7,0.7963)(8,0.8400)(9,0.8726)(10,0.9020)(11,0.9316)(12,0.9524)(13,0.9792)(14,0.9968)(15,1.0127)(16,1.0258)(17,1.0421)(18,1.0441)(19,1.0433)(20,1.0713)};

\nextgroupplot[title={correlation / average},
  ymin=0, ymax=1.35, ytick={0,0.2,0.4,0.6,0.8,1.0,1.2}]
\addplot[mpblue]   coordinates {(1,0.0000)(2,0.1628)(3,0.2264)(4,0.2732)(5,0.3001)(6,0.3224)(7,0.3420)(8,0.3593)(9,0.3741)(10,0.3861)(15,0.4306)(20,0.4624)(25,0.4825)(30,0.4949)(40,0.5103)(50,0.5188)(60,0.5245)(70,0.5263)(80,0.5254)(90,0.5225)(100,0.5154)(110,0.5058)};
\addplot[mporange] coordinates {(1,0.0000)(2,0.3048)(3,0.4771)(4,0.5951)(5,0.6757)(6,0.7408)(7,0.7964)(8,0.8446)(9,0.8856)(10,0.9203)(11,0.9517)(12,0.9802)(13,1.0062)(14,1.0304)(15,1.0532)};
\addplot[mpgreen]  coordinates {(1,0.0000)(2,0.6401)(3,0.9090)(4,1.0816)(5,1.2181)};
\addplot[mpred]    coordinates {(1,0.0000)(2,0.3439)(3,0.4884)(4,0.5811)(5,0.6544)(6,0.6802)(7,0.7068)(8,0.7335)(9,0.7575)(10,0.7828)(11,0.8086)(12,0.8306)(13,0.8547)(14,0.8749)(15,0.8941)(16,0.9102)(17,0.9235)(18,0.9377)(19,0.9484)(20,0.9547)};

%% ════════════════════════════════════════════════════════════
%% ROW 3: euclidean  (ymax ~26)
%% ════════════════════════════════════════════════════════════

\nextgroupplot[title={euclidean / single},
  ylabel={HSE},
  ymin=0, ymax=26, ytick={0,5,10,15,20,25}]
\addplot[mpblue]   coordinates {(1,0.0)(2,4.36)(3,6.17)(4,7.50)(5,8.30)(6,9.03)(7,9.65)(8,10.16)(9,10.61)(10,10.99)(15,12.50)(20,13.43)(25,14.17)(30,14.74)(40,15.56)(50,16.14)(60,16.58)(70,16.87)(80,17.02)(90,17.05)(100,17.02)(110,16.99)};
\addplot[mporange] coordinates {(1,0.0)(2,5.69)(3,8.67)(4,10.43)(5,10.73)(6,11.42)(7,11.99)(8,12.41)(9,12.36)(10,12.04)(11,12.07)(12,11.95)(13,11.81)(14,11.75)(15,11.72)};
\addplot[mpgreen]  coordinates {(1,0.0)(2,13.61)(3,18.12)(4,20.75)(5,22.82)};
\addplot[mpred]    coordinates {(1,0.0)(2,7.31)(3,9.73)(4,11.15)(5,12.26)(6,12.40)(7,12.78)(8,13.53)(9,13.37)(10,13.28)(11,13.41)(12,13.42)(13,13.43)(14,13.34)(15,12.98)(16,12.87)(17,12.74)(18,12.66)(19,12.61)(20,12.43)};

\nextgroupplot[title={euclidean / complete},
  ymin=0, ymax=26, ytick={0,5,10,15,20,25}]
\addplot[mpblue]   coordinates {(1,0.0)(2,4.36)(3,6.49)(4,7.91)(5,9.01)(6,9.93)(7,10.69)(8,11.37)(9,11.94)(10,12.39)(15,14.30)(20,15.58)(25,16.50)(30,17.20)(40,18.31)(50,19.09)(60,19.85)(70,20.37)(80,20.73)(90,20.95)(100,20.93)(110,20.93)};
\addplot[mporange] coordinates {(1,0.0)(2,5.69)(3,8.70)(4,10.71)(5,11.96)(6,12.69)(7,13.23)(8,13.65)(9,13.96)(10,14.14)(11,14.22)(12,14.24)(13,14.22)(14,14.16)(15,14.09)};
\addplot[mpgreen]  coordinates {(1,0.0)(2,13.61)(3,19.90)(4,22.79)(5,24.65)};
\addplot[mpred]    coordinates {(1,0.0)(2,7.31)(3,10.69)(4,12.56)(5,13.96)(6,15.05)(7,15.88)(8,16.39)(9,16.80)(10,17.10)(11,17.30)(12,17.41)(13,17.48)(14,17.56)(15,17.59)(16,17.56)(17,17.49)(18,17.38)(19,17.25)(20,17.11)};

\nextgroupplot[title={euclidean / average},
  ymin=0, ymax=26, ytick={0,5,10,15,20,25}]
\addplot[mpblue]   coordinates {(1,0.0)(2,4.36)(3,6.23)(4,7.58)(5,8.59)(6,9.42)(7,10.08)(8,10.62)(9,11.12)(10,11.56)(15,13.15)(20,14.25)(25,15.11)(30,15.76)(40,16.76)(50,17.48)(60,18.01)(70,18.41)(80,18.75)(90,18.95)(100,19.02)(110,18.96)};
\addplot[mporange] coordinates {(1,0.0)(2,5.69)(3,8.69)(4,10.59)(5,11.46)(6,12.02)(7,12.47)(8,12.85)(9,13.04)(10,13.04)(11,13.00)(12,12.92)(13,12.80)(14,12.67)(15,12.56)};
\addplot[mpgreen]  coordinates {(1,0.0)(2,13.61)(3,19.01)(4,21.55)(5,23.35)};
\addplot[mpred]    coordinates {(1,0.0)(2,7.31)(3,10.21)(4,11.58)(5,12.55)(6,13.38)(7,14.00)(8,14.64)(9,14.99)(10,15.06)(11,15.13)(12,15.19)(13,15.18)(14,15.15)(15,15.02)(16,14.94)(17,14.79)(18,14.63)(19,14.50)(20,14.34)};

%% ════════════════════════════════════════════════════════════
%% ROW 4: manhattan  (ymax ~680)
%% ════════════════════════════════════════════════════════════

\nextgroupplot[title={manhattan / single},
  ylabel={HSE},
  ymin=0, ymax=680, ytick={0,100,200,300,400,500}]
\addplot[mpblue]   coordinates {(1,0)(2,130)(3,165)(4,193)(5,205)(6,216)(7,221)(8,226)(9,234)(10,238)(15,250)(20,263)(25,280)(30,285)(40,304)(50,309)(60,309)(70,317)(80,314)(90,308)(100,301)(110,296)};
\addplot[mporange] coordinates {(1,0)(2,126)(3,186)(4,213)(5,194)(6,199)(7,203)(8,204)(9,192)(10,176)(11,171)(12,163)(13,156)(14,151)(15,148)};
\addplot[mpgreen]  coordinates {(1,0)(2,430)(3,481)(4,507)(5,533)};
\addplot[mpred]    coordinates {(1,0)(2,231)(3,258)(4,273)(5,287)(6,268)(7,259)(8,271)(9,246)(10,232)(11,228)(12,223)(13,217)(14,210)(15,195)(16,189)(17,183)(18,178)(19,175)(20,167)};

\nextgroupplot[title={manhattan / complete},
  ymin=0, ymax=680, ytick={0,100,200,300,400,500,600}]
\addplot[mpblue]   coordinates {(1,0)(2,130)(3,182)(4,215)(5,241)(6,263)(7,281)(8,297)(9,310)(10,319)(15,355)(20,386)(25,404)(30,417)(40,443)(50,458)(60,468)(70,473)(80,474)(90,466)(100,471)(110,472)};
\addplot[mporange] coordinates {(1,0)(2,126)(3,187)(4,225)(5,243)(6,248)(7,250)(8,251)(9,250)(10,247)(11,243)(12,237)(13,232)(14,226)(15,221)};
\addplot[mpgreen]  coordinates {(1,0)(2,430)(3,586)(4,624)(5,638)};
\addplot[mpred]    coordinates {(1,0)(2,231)(3,315)(4,352)(5,378)(6,397)(7,408)(8,410)(9,411)(10,407)(11,402)(12,396)(13,392)(14,387)(15,381)(16,375)(17,367)(18,359)(19,352)(20,344)};

\nextgroupplot[title={manhattan / average},
  ymin=0, ymax=680, ytick={0,100,200,300,400,500}]
\addplot[mpblue]   coordinates {(1,0)(2,130)(3,169)(4,197)(5,218)(6,235)(7,247)(8,258)(9,268)(10,276)(15,304)(20,325)(25,338)(30,348)(40,363)(50,373)(60,379)(70,381)(80,381)(90,383)(100,378)(110,374)};
\addplot[mporange] coordinates {(1,0)(2,126)(3,186)(4,220)(5,223)(6,222)(7,221)(8,220)(9,216)(10,208)(11,201)(12,193)(13,185)(14,178)(15,172)};
\addplot[mpgreen]  coordinates {(1,0)(2,430)(3,533)(4,552)(5,563)};
\addplot[mpred]    coordinates {(1,0)(2,231)(3,287)(4,297)(5,303)(6,308)(7,311)(8,317)(9,318)(10,307)(11,299)(12,293)(13,286)(14,280)(15,270)(16,263)(17,254)(18,246)(19,238)(20,233)};

\end{groupplot}

%% Shared legend below all panels
\node[anchor=north] at
  ($(group c1r4.south east)!0.5!(group c3r4.south west) - (0,0.8cm)$)
  {\pgfplotslegendfromname{sweeplegend}};

\end{tikzpicture}
\caption{Normalised HSE as a function of society size across all combinations
  of distance metric (rows: cosine, correlation, euclidean, manhattan) and
  linkage method (columns: single, complete, average), using greedy agent
  selection on EmbedLLM. Cosine and correlation panels are normalised by the
  maximum achievable HSE; euclidean and manhattan panels show raw HSE values.
  The qualitative pattern --- specialist societies plateau rapidly at high HSE
  while the default real-world pool plateaus at a substantially lower level ---
  is consistent across all twelve configurations.}
\label{fig:hse_param_sweep}
\end{figure*}

\section{Effect of Distance Metric and Linkage on HSE}
\label{app:linkage}

Figure~\ref{fig:hse_param_sweep} reports normalised HSE as a function
of society size across all twelve combinations of distance metric
(cosine, correlation, euclidean, manhattan) and linkage method
(single, complete, average), using greedy agent selection on EmbedLLM.
The qualitative pattern is consistent across configurations: specialist
societies plateau rapidly at high HSE while the default real-world pool
plateaus at a substantially lower level, confirming that the finding
reported in Section~\ref{sec:results} does not depend on the choice of
distance or linkage. Cosine distance with single linkage, used throughout
the main paper, produces the most conservative estimates of diversity
for the default society.%, making it the most cautious choice for
%supporting our main claims.

\section{Effect of Model Selection Criteria}
\label{sec:model:selection}

\newcolumntype{R}{>{\raggedleft\arraybackslash}X}

\begin{table}[t]
% \begingroup
\small 
\setlength{\tabcolsep}{2pt} 
\begin{tabularx}{\linewidth}{l r *{6}{R}}
\toprule
& $n$ & HSE & rand. & KNN1 & KNN3 & KNN10 & orac.\\
\midrule
 default & 112 & 0.24 & 0.43 & 0.52 & 0.57 & 0.65 & 0.97 \\
 \midrule
\multirow{3}{*}{HSE} & 9 & 0.53 & 0.30 & 0.34 & 0.40 & 0.46 & 0.87 \\
 & 12 & 0.49 & 0.30 & 0.35 & 0.41 & 0.46 & 0.89 \\
 & 25 & 0.41 & 0.35 & 0.48 & 0.50 & 0.52 & 0.94 \\
 \midrule
\multirow{3}{*}{HSE + acc.} & 9 & 0.48 & 0.33 & 0.52 & 0.59 & 0.65 & 0.92 \\
 & 12 & 0.47 & 0.34 & 0.52 & 0.58 & 0.65 & 0.93 \\
 & 25 & 0.41 & 0.35 & 0.46 & 0.55 & 0.63 & 0.96 \\
\bottomrule
\end{tabularx}
% \endgroup
\caption{Task accuracy and society diversity (normalised HSE) for different routing policies over subsets of the EmbedLLM society selected 
under two criteria: maximising HSE alone, and 
maximising HSE jointly with task accuracy.}
\label{tab:joint:criterion}
\end{table}
% \begin{tabular}{lrrrrrr}
% \toprule
% criteria & n & hse & random & KNN 1 & KNN 10 & oracle \\
% \midrule
%  & 112 & 1.62 & 0.43 & 0.52 & 0.65 & 0.97 \\
% \multirow{3}{*}{hse} & 9 & 1.67 & 0.30 & 0.34 & 0.46 & 0.87 \\
%  & 12 & 1.76 & 0.30 & 0.35 & 0.46 & 0.89 \\
%  & 25 & 1.91 & 0.35 & 0.48 & 0.52 & 0.94 \\
% \multirow{3}{*}{hse + score} & 9 & 1.54 & 0.33 & 0.52 & 0.65 & 0.92 \\
%  & 12 & 1.68 & 0.34 & 0.52 & 0.65 & 0.93 \\
%  & 25 & 1.88 & 0.35 & 0.46 & 0.63 & 0.96 \\
% \bottomrule
% \end{tabular}

Table~\ref{tab:joint:criterion} reports task accuracy and society diversity for subsets of the EmbedLLM society selected under two criteria: maximising HSE alone, and maximising HSE jointly with task accuracy. The results reveal a diversity-accuracy tradeoff. Subsets selected purely on HSE achieve higher normalised HSE at every size but lower routing task accuracy. Adding task accuracy to the selection criterion recovers routing performance --- at $n{=9}$, KNN-10 matches the full 112-model society exactly (0.65) and oracle reaches 0.92 --- but at the cost of lower HSE (0.48 vs.\ 0.53). The HSE\,+\,acc. subset at $n{=9}$ achieves competitive routing accuracy using fewer that 10\% of the available models, providing a practical coreset that balances diversity and performance.